\pdfoutput=1
\documentclass[
    aps,
    prd,
    onecolumn,
    superscriptaddress,
    nofootinbib,
    amsmath,amssymb,
]{revtex4-2}

\PassOptionsToPackage{table}{xcolor}

\makeatletter\let\switch@array\relax\makeatother

\usepackage[T1]{fontenc}
\usepackage[utf8]{inputenc}
\usepackage[english]{babel}
\usepackage{microtype}

\usepackage{mathtools}   %
\usepackage{amssymb}
\usepackage{bm}

\newcommand{\op}[1]{\textsc{#1}}

\newcommand{\cottax}{\texttt{cottax}}

\usepackage{graphicx}
\graphicspath{{figures/}}
\usepackage{booktabs}
\usepackage{siunitx}
\usepackage{subcaption}
\usepackage{newfloat}
\DeclareFloatingEnvironment[name=Algorithm,fileext=loa,listname={List of Algorithms}]{algorithm}
\usepackage{algpseudocode}
\usepackage{tikz}
\usetikzlibrary{arrows.meta, shapes.geometric, shapes.symbols, positioning, fit, calc, patterns.meta}
\usepackage{pgfplots}
\pgfplotsset{compat=1.18}

\usepackage{listings}
\usepackage{xcolor}
\definecolor{c_grey_light}{RGB}{223,223,223}
\definecolor{lstbg}{HTML}{F7F7F4}
\definecolor{lstkw}{HTML}{75588C}
\definecolor{lstcm}{HTML}{6F7B87}
\definecolor{lststr}{HTML}{A8742A}
\usepackage{enumitem}
\newlist{checks}{enumerate}{1}
\setlist[checks]{label=(\thesection.\arabic*), ref=(\thesection.\arabic*), leftmargin=*, resume=checks}

\newcommand{\incomp}{\mathrel{\parallel}}    %
\newcommand{\defiff}{\mathrel{\vcentcolon\Leftrightarrow}}

\makeatletter
\AtBeginDocument{%
  \typeout{FIGLAYOUT columnwidth=\the\columnwidth\space textwidth=\the\textwidth\space fontsize=\f@size}%
}
\makeatother

\usepackage{hyperref}
\hypersetup{
    colorlinks=true,
    linkcolor=blue!50!black,
    citecolor=blue!50!black,
    urlcolor=blue,
}
\usepackage[noabbrev,capitalise]{cleveref}
\crefname{checksi}{Check}{Checks}
\Crefname{checksi}{Check}{Checks}
\newtheorem{gcheck}{Check}

\newtheorem{echeck}{Check}

\crefname{gcheck}{check}{checks}
\Crefname{gcheck}{Check}{Checks}
\crefname{echeck}{check}{checks}
\Crefname{echeck}{Check}{Checks}
\crefname{requirement}{requirement}{requirements}
\Crefname{requirement}{Requirement}{Requirements}

\begin{document}

\title{Managing complexity for integrated fusion reactor design using a differentiable graph-based MDAO approach: PROCESS as a case study}

\author{Timo J. Bogaarts}
\email{timobogaarts@gmail.com}
\affiliation{Eindhoven University of Technology, PO Box 513, 5600 MB, Eindhoven, Netherlands}

\author{Willem Rutten}
\affiliation{Eindhoven University of Technology, PO Box 513, 5600 MB, Eindhoven, Netherlands}

\author{Felix Warmer}
\affiliation{Eindhoven University of Technology, PO Box 513, 5600 MB, Eindhoven, Netherlands}
\affiliation{Max-Planck-Institut für Plasmaphysik, Teilinstitut Greifswald, Wendelsteinstrasse 1, Greifswald D-17491, Germany}

\date{\today}

\begin{abstract}
Integrated design of nuclear fusion reactors requires combining an increasingly large number of models from different disciplines self-consistently, with model selection and parametrisation changing between design studies. 
Manual definition of execution order and solvers for cyclic dependencies is error-prone, inefficient, and inflexible.
We propose to use a generic graph-based coordination and execution core that enables automated execution order, cycle detection, and solver architecture specification for any set of models or parametrisation.
To this end, a computational dependency graph of purely functional models is extended by a conditional node and nesting edges, treating solvers and optimisers in the same way.
A minimal set of graph operations then allows for fine-tuned control on feedback resolution, making both optimiser and solver architectures easily reconfigurable.
Existing monolithic multidisciplinary design analysis and optimisation (MDAO) architectures can be recovered by sequentially applying these operations.
The \cottax{} package implements this extended dependency graph, graph operations, executability checks, and a JAX-based execution layer. 
This execution layer eliminates runtime overhead and provides native gradient computation through the whole graph.
As a proof of principle, \cottax{} is then applied to the PROCESS systems code. 
Graph sequencing shows that for PROCESS only three small coupled components need to be solved every optimiser iteration. 
Monolithic MDAO architectures are then assigned to seven different PROCESS configurations, demonstrating the ease of (re)configuring its architecture. 
Compared to PROCESS, the sequencing and execution back-end together yield a two to three orders of magnitude reduction in model runtime during optimisation.
\cottax{} is currently being considered as the basis for developing an integrated stellarator MDAO framework.
\end{abstract}  

\maketitle

\section{Introduction}\label{sec:introduction}
Nuclear fusion reactors are complex to design, requiring highly specialised teams of experts working on interdependent subsystems. 
A complete engineering design may take years, which precludes the many evaluations required for design space exploration and optimisation. 
In practice, exploration and optimisation are thus carried out with a reduced parametrisation of the reactor (a small set of high-level decisions such as major radius, magnetic field strength, net electric power, blanket type and thickness) and with reduced-fidelity models of its behaviour \cite{federici2017european}.

Fusion systems codes \cite{kovari2016process,kovari2014process,reux2015demo} integrate a set of low- or medium-fidelity models self-consistently to compute system-level design objectives (e.g. cost of electricity) and engineering constraint satisfaction as a function of these high-level decisions. 
With an optimisation algorithm~\cite{nocedal_numerical_2006} these models are then executed iteratively to find an optimal design candidate within this parametrisation --- an approach more generally known as multidisciplinary design analysis and optimisation (MDAO) \cite{sobieszczanski1997multidisciplinary,mdobook,gray2019openmdao}. 
The resulting optimum serves as the starting point for detailed engineering design.

However, the behaviour predicted by the low-fidelity models often turns out to be inconsistent with that obtained in detailed engineering design \cite{coleman2019blueprint,franza2019development} (e.g. the breeding blanket might not produce its target tritium breeding ratio when 3D effects are taken into account). 
Since the optimality and engineering constraint satisfaction of the design candidate were established with respect to the low-fidelity models, such inconsistencies imply it is at best non-optimal and at worst unviable, in which case the high-level decisions must change. 
This is expensive: the strong coupling between subsystems \cite{federici2017european} means such a change rarely stays local.
For example, a thicker blanket also affects the coil geometry, vacuum vessel, and support structure, which then all require a detailed redesign. 

Using coupled higher-fidelity parametrisations and models directly in the system-level analysis and optimisation reduces the chance of such inconsistencies and thereby the need for detailed design iterations \cite{coleman2019blueprint,meneghini2024fuse,franza2022mira}. Higher-fidelity system-level analysis and parametrisation is especially crucial for stellarators. 
First, their three-dimensional shaping increases the likelihood of inconsistencies arising between analytic models or scaling laws (typically adapted from tokamaks \cite{warmer2015implementation,lion2021general}) and a detailed design. 
Second, this shaping introduces additional design parameters whose effects low-fidelity models cannot capture. 
Finally, the plasma shape is typically optimised first --- before the system-level analysis --- using low-fidelity proxies for the engineering design (e.g. required plasma-coil distance), which again makes non-optimal or unviable designs more likely.

\begin{figure}
    \centering
    \input{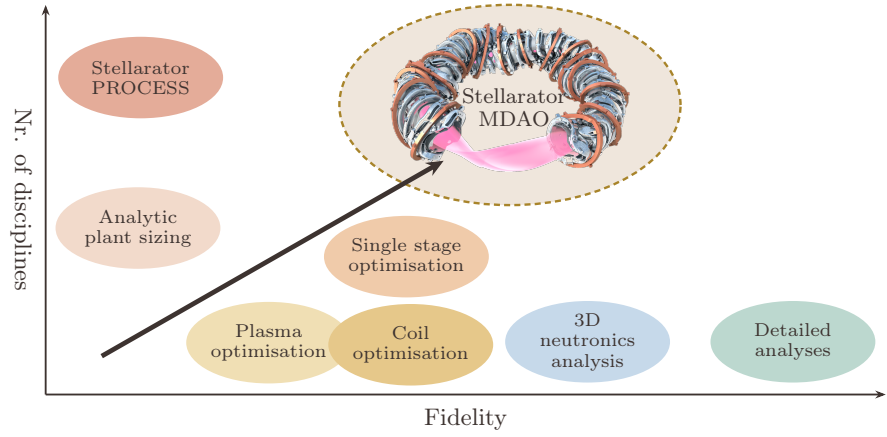} %
    \caption{Stellarator reactors are currently analysed either with high fidelity or with a large number of disciplines. Stellarator MDAO aims to combine higher-fidelity models and a larger number of disciplines.}
    \label{fig:intro}
\end{figure}

Yet, as illustrated in \cref{fig:intro}, current stellarator analyses are either low-fidelity system-level analyses \cite{warmer2015implementation,lion2021general} or high-fidelity few-discipline codes that cannot be readily coupled. Within the Stellarator-Reactor-Studies group at IPP Greifswald, a stellarator MDAO framework is being developed to fill this gap~\cite{warmer2024overview}, enabling self-consistent, higher-fidelity stellarator design space exploration and optimisation and thereby accelerating the design process. 

No single set of models and parametrisations (or even a common data structure) has been established for higher-fidelity stellarator MDAO. Moreover, the choice of parametrisation and model set could change per design study: for example, surrogates for optimisation around a known design point, low-fidelity models for large-scale design space exploration, or mixed-fidelity models to balance computational cost and accuracy. Even with a fixed model set and parametrisation, the problem formulation (which variables can vary, which constraints are considered, and which objective is pursued) may change: e.g. minimising major radius rather than maximising net electric power. Fixed choices for how models are coordinated and executed are thus incompatible with the development and usage of such a framework. Therefore, we argue for a generic core that handles the coordination and execution in a design-problem-agnostic way. 

\subsection{Requirements for the coordination and execution core}\label{sec:requirements}
The coordination and execution core of the stellarator MDAO framework needs to satisfy a set of requirements.

First, as argued, the core should support arbitrary design problems: changing the models, parametrisation, or problem formulation should not require changing the core itself.

Second, the core should be able to automatically generate a valid and efficient execution order of the models and their solvers (an MDAO architecture) for such arbitrary design problems. Manual coordination is feasible for a small set of models, but becomes error-prone and time-intensive as the number of models and their interdependencies increase. Furthermore, it has to be repeated for every change in model set, parametrisation, or problem formulation. A fixed-point iteration over all models to resolve coupling is always possible, but is not efficient (uncoupled models would be run unnecessarily) and may not converge at all.

Third, an automatically assigned architecture is not always the most efficient~\cite{martins2013multidisciplinary}. For example, the vastly different computational costs of different fidelity levels might warrant nesting of solvers or optimisers to iterate the lower-fidelity models more often (similar to the motivation for Asymmetric Subspace Optimization \cite{chittickAsymmetricSuboptimizationApproach2009,mdobook}). Therefore, the core should be able to implement arbitrary solver and optimisation structures and allow for easy modifications whenever required by the particular problem (which, again, changes with the design problem).

Fourth, higher-fidelity parametrisations (e.g. plasma boundary or blanket shape parameters) have significantly more design variables than low-fidelity ones (e.g. size or average field strength). Gradient-free optimisation does not scale well in this regime \cite{mdobook}, and therefore gradient-based methods should be natively supported and integrated. 
To minimise implementation effort and avoid the scaling and accuracy issues of finite differencing, automatic differentiation methods and implicit differentiation (e.g. adjoint methods) are strongly preferred \cite{hwang2019large}. 

Finally, the core should not dominate runtime even when extremely fast, low-fidelity models are used.
For applications such as uncertainty quantification \cite{lux2017uncertainties} or design database generation \cite{giuliani2025comprehensive,cadena2026constellaration}, fast design analysis can be crucial. 

Together this yields the following set of requirements:

\begin{enumerate}
    \item\label[requirement]{req:selection} \textbf{Design problem agnostic}: any design problem (a selection of models, their parametrisation, and a problem formulation) should be usable without changes to the core.
    \item\label[requirement]{req:coordination} \textbf{Automated coordination}: the execution order and solver structure of models should be automatically derivable for any such design problem.
    \item\label[requirement]{req:flexible} \textbf{Flexible MDAO architecture}: the MDAO architecture should be easily modifiable beyond a default implementation to allow (re)configuration of analysis or optimisation problems. %
    \item\label[requirement]{req:gradients} \textbf{Gradient computation}: gradients should be available natively using automatic and implicit differentiation.
    \item\label[requirement]{req:overhead} \textbf{Low overhead}: the core itself should not dominate runtime even for low-fidelity fast models. %
\end{enumerate}

Building on existing state-of-the-art fusion design codes or general-purpose MDAO frameworks would be preferable; however, this combination of requirements defines a particular niche that falls outside their scope.
In fusion, design optimisation codes such as PROCESS \cite{kovari2014process,kovari2016process}, BLUEMIRA \cite{bluemira,coleman2019blueprint,franza2022mira} or SIMSOPT \cite{landreman2021simsopt} have a modular software architecture, but require manual specification of model execution order and only support a hard-coded MDAO architecture, making them unsuitable for building the stellarator MDAO framework on.
Recent codes such as DESC \cite{dudt2020desc} and FUSE \cite{meneghini2024fuse} can use advanced gradient computation (JAX and Julia, respectively) but also lack the automated coordination we require.

Outside of fusion, general-purpose multidisciplinary design and optimisation frameworks such as OpenMDAO \cite{gray2019openmdao} and GEMSEO \cite{gemseo_paper} in principle provide flexibility in MDAO architecture, but either require significant manual coordination to reconfigure MDAO architectures (OpenMDAO) or are limited in flexibility of the MDAO architecture (GEMSEO).
Novel approaches such as the AGILE paradigm \cite{van2019agile, ciampaAGILEParadigmNext2020} use a graph-based approach to automate the coordination and ease (re)configuration for predefined optimisation architectures, but the openly available execution layer introduces too much overhead for our purposes. 
For a more detailed discussion, see \cref{sec:literature}.

This has motivated the development of the herein presented graph-based generic coordination and execution core \cottax{} that aims to satisfy these requirements.

\subsection{Contribution}
Similar to \cite{pate2014graph,van2019agile,gemseo_paper}, we use a graph approach based on purely functional models connected by their input/output variables.
The resulting graph allows the use of topological ordering and strongly connected component algorithms \cite{tarjanDepthFirstSearchLinear1972} that find the smallest groups of coupled models that have to be solved together.
Whereas these previous approaches start from the definition of an MDAO problem and then derive an architecture from it, we start from the definition of what an executable graph is and allow any graph that passes a set of executability checks to be executed (whether it corresponds to a standard MDAO architecture or not). 

The contribution of this work is then twofold: by enabling the graph to represent any MDAO architecture, we extend the flexibility of a graph-based approach, and we allow this same graph to be directly and efficiently executed using JAX.

Our first contribution is the extension of a computational dependency graph by \emph{conditional nodes} and \emph{nesting edges} supported with a set of mathematical checks that define when this extended graph is executable.
These conditional nodes are a generalisation of optimiser and solver iteration loops that can represent (among others) a fixed-point iteration, a residual (root-finding) solver, an optimisation problem, a consistency constraint, and an ordinary differential equation.
Nesting edges indicate which conditional node iterates which other conditional node.
The executability checks are generic: they only depend on the conditions and unknowns of a conditional node, and not on its type. %
The unification of solvers and optimisers as a single conditional node type combined with the nesting allows the creation of highly customised MDAO architectures: solvers and optimisers can be freely nested and combined with each other. The optimiser, solvers converging coupling variables, and the solvers of individual models have no special role: they are conditional nodes like any other.
We also define a small set of graph operations that can turn any design problem into an executable graph. Standard monolithic MDAO architectures are then a sequential application of these graph operations instead of predefined presets, and can therefore be automatically assigned and easily modified. 
This graph representation, the executability checks, the graph operations, and the architectures are implemented in the Python package \cottax{} in a generic way (i.e. an MDAO architecture can be defined and checked without implementations of the models and solvers). %

Our second contribution is the execution of this extended graph using JAX \cite{jax2018github}: the entire \cottax{} graph, including its nested solves and its orchestration, becomes one transformable JAX function, enabling efficient execution, differentiation, and batching. The JAX just-in-time compilation then removes the \cottax{} machinery entirely, producing a program equivalent to a manual ordering and ensuring that the runtime is spent entirely within the models and their (possibly coupled) solves. 
JAX's automatic differentiation can be used over the entire dependency graph, including over (custom) solvers using implicit differentiation. 
Non-JAX functions (e.g. legacy models) can be wrapped, with finite differences as a backup when derivatives (from automatic differentiation) are not available (e.g. a meshing step). 

To demonstrate \cottax{}, we apply it to the PROCESS fusion systems code \cite{kovari2014process,kovari2016process}. PROCESS has over 140 different models with complex interdependencies and has different configurations that share less than half of their models, have different parametrisations, and have different problem formulations. Therefore, it provides a compelling test case for \cref{req:selection,req:coordination}. Since the models themselves are mostly fast scalar arithmetic, it also provides a test for the execution speed of the \cottax{} graph machinery. 
First, we apply the automated coordination to seven different PROCESS configurations, showing how multiple MDAO architectures can be automatically assigned and run in all configurations, addressing \cref{req:selection,req:coordination}. Moreover, the results motivate \cref{req:flexible}: no single MDAO architecture is best in all cases.
Second, to investigate \cref{req:gradients,req:overhead}, we profile the \cottax{} JAX-based execution for these seven configurations. Automatic and implicit differentiation (including through nested solves) is used for the required Jacobians in all configurations. The timing results show two to three orders of magnitude less model time (i.e. not including driver computations) in \cottax{} compared to base PROCESS, and show that removing orchestration overhead (by compiling the \cottax{} graph) results in a $40\times$ speedup in a stellarator configuration.

The remainder of this paper is organised as follows. In \cref{sec:literature}, the state of the art of model coordination is given. 
Then, in \cref{sec:methods}, the dependency graph, visualisations of it, conditional nodes and nesting edges, the workflow for getting to an executable graph, and automated MDAO architecture recipes are introduced.
The \cottax{} implementation is discussed in \cref{sec:imp}.
In \cref{sec:process}, we apply it to the PROCESS systems code~\cite{kovari2014process,kovari2016process}, obtaining the dependency graph, sequencing it, and solving it with varying architectures.
Finally, a discussion and conclusions are given in \cref{sec:discussion,sec:conclusions}.

\section{State of the art}
\label{sec:literature}

Complex dependencies are often modelled as graphs, giving rise to classic systems-engineering diagrams such as the $N^2$~\cite{lanoN2Chart1977, buedeEngineeringDesignSystems2016} or Design Structure Matrix \cite{stewardDesignStructureSystem1981, eppingerDesignStructureMatrix2012}, allowing for analysis such as clustering or minimal cycles.  
The dependencies between models in integrated design codes can be similarly encoded and visualised as a graph.
Typically, this is done with an extended Design Structure Matrix that reveals the computational structure~\cite{lambe2012extensions}.
In fact, depending on the implementation approach such a dependency graph can be directly derived from the software~\cite{ma2021modelingtoolkit,gray2019openmdao}.

Aside from analysis of existing codes, graphs can also be used to formulate MDAO problems~\cite{pate2014graph}.
This allows sequencing algorithms such as Tarjan~\cite{tarjanDepthFirstSearchLinear1972} to be used to automatically create a model order and isolate so-called strongly connected components. 
In essence, this is a reordering of computational tasks that can be related to computational optimisation approaches as used in compilers, where computational graphs allow the program to be aggressively optimised~\cite{xla}.
For example, the statically known graph enables efficient ordering and dead-code elimination.
Similar strategies have been applied to dependency graphs at a higher abstraction level such as individual disciplines in an MDAO context~\cite{rogersDeMAIDDesignManagers1989,browningApplyingDesignStructure2001, deborstDecompositionAnalysisMultidisciplinary2016}. %
Several state-of-the-art frameworks integrate such approaches in different ways.

OpenMDAO \cite{gray2019openmdao, openmdao3.45} is an established multidisciplinary framework, providing efficient derivatives of large-scale analysis using the MAUD architecture \cite{Hwang2018a}. 
It focuses on monolithic architectures that may contain nested solvers, but does not provide the desired flexibility to easily (re)configure a large collection of models. 
It can automatically find model execution order within a model Group, but does not facilitate easy (re)configuration of those Groups when cyclic model dependencies are changing.
Reconfiguration of the MDAO architecture requires manually recomposing the Groups.
OpenMDAO supports MPI~\cite{mpi4py} and PETSc~\cite{petsc4py} to parallelise model groups and large linear solvers (such as required for MAUD Jacobian computation), and JAX-based components can be included; however, the orchestration layer remains in Python.
We prefer to avoid the overhead with a JAX-native orchestration layer for cases with many small, fast models.

GEMSEO~\cite{gemseo_paper, gemseo6.3} is a Python-based engine for multidisciplinary design exploration and optimisation that supports both distributed and monolithic architectures.
Though GEMSEO can automatically order models and assign a preconfigured MDAO architecture, it does not support easily modifying which solver converges which variables and models. 
Similar to OpenMDAO, the nesting in the MDAO architecture needs to be hard-coded for the specific application.
In terms of gradient handling, GEMSEO allows both direct and adjoint propagation using coupled adjoint theory~\cite{martins2013review}.
With a plugin~\cite{gemseo-jax} (a chain of) JAX-based disciplines can be supported, but the framework itself is not JAX-native so some overhead remains.

The AGILE project \cite{ciampaAGILEParadigmNext2020, van2019agile,boggeroAGILE40MBSEMDAO2022} extensively uses graphs to configure optimisation workflows with the KADMOS package \cite{van2019formulation}, building on prior work of Pate et al.~\cite{pate2014graph}.
Part of this (re)configuration support includes algorithmic ordering and partitioning of the graph, and KADMOS provides preset architectures to implement.
Custom architectures that e.g. assign couplings to specific drivers or have deep nesting require manually defining new architecture presets.
AGILE defines couplings to various execution layers (incl. RCE~\cite{bodenRCEIntegrationEnvironment2021} and OpenMDAO~\cite{gray2019openmdao} via OpenLEGO~\cite{devriesIndustrializationMDAOEvolving2017}), but focuses more on multi-institutional collaborative infrastructure~\cite{moerlandCollaborativeArchitectureSupporting2020}. 
These implementations therefore are not particularly suited for low-latency coupling of autodifferentiable models.

AeroSandbox~\cite{sharpe2021aerosandbox,sharpe2024accelerating} similarly considers a computational graph and transformations thereof, but argues for a procedural programming interface --- on the grounds that a functional programming style requires careful ordering, significant boilerplate, and inflexible problem and constraint definitions during scripting. 
However, as the above references (\cite{van2019agile,gemseo_paper}) have shown, topological sorting can eliminate the first objection.
As \cite{sharpe2024accelerating} notes, modular disciplines are increasingly attractive for high-fidelity and complex optimisation problems, such as we consider in this work.
Furthermore, the adoption of a JAX-based approach removes any boilerplate-induced computational overhead.
Lastly, a functional programming style maps naturally to a graph-based approach that precisely allows high flexibility in problem formulation, as shown by \cite{van2019formulation}.

Beyond MDAO frameworks, (systems-level acausal) modelling languages such as Modelica \cite{mattsson1998physical} or ModelingToolkit.jl \cite{ma2021modelingtoolkit} similarly build a computational graph from equations, ordering automatically and resolving the induced functional cycles.
However, the frameworks have different use-cases, often with a lower granularity (symbolic scalar equations instead of array-based opaque models), and because they are not geared towards (multidisciplinary) optimisation are not easily adapted to fusion design optimisation problems.

Fusion (integrated) design approaches typically feature custom frameworks.
For example, PROCESS \cite{kovari2014process,kovari2016process} uses a fixed model order with hand-picked iteration variables and consistency constraints; FUSE \cite{meneghini2024fuse}, BLUEMIRA \cite{bluemira,coleman2019blueprint,franza2022mira}, and SYCOMORE \cite{reux2015demo} also hand-declare their analysis workflows.
SIMSOPT \cite{landreman2021simsopt} and DESC~\cite{dudt2020desc} are frameworks for stellarator optimisation which also require explicit manual ordering and MDAO architecture specification.

\begin{table}[htbp]
    \centering\small
    \caption{An overview of the open-source MDAO frameworks that have been considered as the basis for a future stellarator reactor design framework.
    }
    \label{tab:sota_overview}
    \newcommand{\tyes}{$\checkmark$}
\newcommand{\tpart}{$\sim$}
\newcommand{\tno}{--}
\newcommand{\tnote}[1]{{\enspace\scriptsize #1}}
\rowcolors{2}{}{c_grey_light}
\setlength{\extrarowheight}{2pt}
\setlength{\tabcolsep}{0pt}
\begin{tabular}{>{\begin{minipage}[c]{\dimexpr0.27\linewidth+4pt}\vspace{3pt}\raggedright\arraybackslash\rightskip=4pt plus 1fil}l<{\vspace{3pt}\end{minipage}}*{5}{>{\begin{minipage}[c]{\dimexpr(0.73\linewidth-4pt)/5}\vspace{3pt}\centering\arraybackslash\leftskip=4pt plus 1fil\rightskip=4pt plus 1fil}c<{\vspace{3pt}\end{minipage}}}}
    \toprule
    & Design problem agnostic & Automated coordination & Flexible MDAO architecture & Gradient computation & Low overhead \\[2pt]
    \noalign{\xdef\sotabelowrulesep{\the\belowrulesep}\global\belowrulesep=0pt}
    \midrule
    \noalign{\global\belowrulesep=\sotabelowrulesep}
    OpenMDAO~\cite{gray2019openmdao}
        & \tyes
        & \tpart \tnote{within model Group}
        & \tpart \tnote{any nesting, manual}
        & \tyes \tnote{MAUD}
        & \tpart \tnote{Python orchestration} \\[2pt]
    GEMSEO~\cite{gemseo_paper}
        & \tyes
        & \tyes
        & \tpart \tnote{presets}
        & \tyes \tnote{coupled adjoint}
        & \tpart \tnote{Python orchestration} \\[2pt]
    AGILE / KADMOS~\cite{ciampaAGILEParadigmNext2020,van2019formulation}
        & \tyes
        & \tyes
        & \tpart \tnote{presets}
        & \tpart \tnote{via exec. layer}
        & \tno \tnote{file-based} \\[2pt]
    AeroSandbox~\cite{sharpe2024accelerating}
        & \tpart \tnote{procedural script}
        & \tno
        & \tno \tnote{SAND only}
        & \tyes \tnote{AD}
        & \tyes \\[2pt]
    PROCESS~\cite{kovari2014process,kovari2016process}
        & \tpart \tnote{modular code}
        & \tno
        & \tno
        & \tpart \tnote{FD}
        & \tpart \tnote{Python orchestration} \\[2pt]
    BLUEMIRA~\cite{bluemira}, SIMSOPT~\cite{landreman2021simsopt}
        & \tpart \tnote{modular code}
        & \tno
        & \tno
        & \tpart \tnote{FD, partially analytic}
        & \tpart \tnote{Python orchestration} \\[2pt]
    DESC~\cite{dudt2020desc}, FUSE~\cite{meneghini2024fuse}
        & \tpart \tnote{modular code}
        & \tno
        & \tno
        & \tyes \tnote{JAX / Julia AD}
        & \tyes \\[2pt]
\end{tabular}

\end{table}

Overall, various MDAO frameworks have been considered as a basis for a stellarator design framework.
As \cref{tab:sota_overview} summarises, the available frameworks do not sufficiently fulfil the needs for our application.
This has motivated the development of \cottax{}, which will now be introduced.

\clearpage
\section{Theory for graph-based coordination and execution}\label{sec:methods} %
This section describes how the graph-based approach of \cottax{} can turn a set of functions or models into an executable analysis that can be reconfigured with simple graph operations (\cref{req:selection,req:coordination,req:flexible}). First, well-known material concerning pure functions, dependency graphs, orderings, and visualisations of such graphs is covered in \cref{sec:pure_functions}. Then, the extensions of the dependency graph by conditional nodes and nesting edges are discussed in \cref{sec:cycle_resolution,sec:nest_combine}. The checks an executable graph must pass are then given in \cref{sec:workflow}. Finally, \cref{sec:architectures} shows how monolithic MDAO architectures can be obtained from a set of simple graph operations.

Mathematical detail is kept to a minimum; formal definitions of both the node types and the executability criteria are instead given in \cref{app:maths} together with a comparison to similar graph-based MDAO approaches \cite{bruggeman2026dynamic,van2019agile}.

\subsection{Pure functions and dependency graphs}\label{sec:pure_functions}
Pure functions are referentially transparent functions without any side effects: they do not mutate or read global state, they do not mutate their inputs, do not write to files or to the terminal, and only return an output that is solely dependent on its inputs.
The execution of a function therefore cannot affect the execution of another function unless through its inputs.

Given two functions $N_{i}$ and $N_j$, we can therefore reason that $N_i$ only depends on $N_j$ if its inputs $I_i$ contain an element of the outputs $O_j$ of $N_j$.
If not, $N_i$ is free to execute before $N_j$, and moreover, if $N_j$ conversely does not depend on $N_i$, they can be executed in parallel.
Such a dependency structure can be formalised in a computational dependency graph (hereafter, dependency graph) with the pure functions as nodes, connected by directed edges labelled by the variables.

\begin{figure}[b]
    \centering
    \begin{subfigure}[c]{0.26\linewidth}
        \centering
        \includegraphics[width=\linewidth]{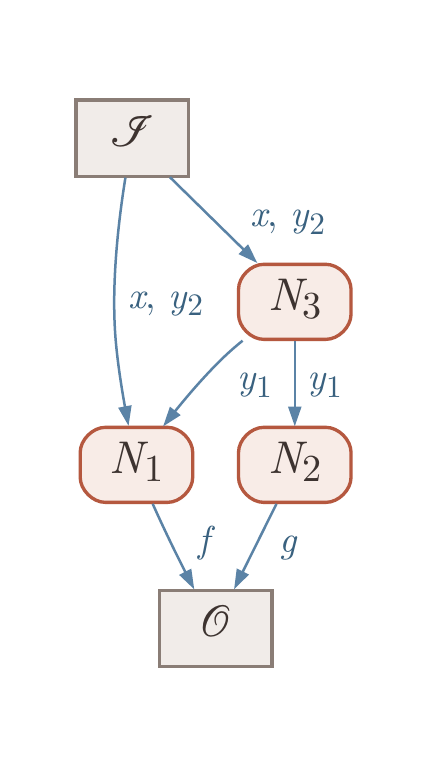}
        \caption{}\label{fig:basic_graph}
    \end{subfigure}
    \hfill
    \begin{subfigure}[c]{0.34\linewidth}
        \centering
        \includegraphics[width=\linewidth]{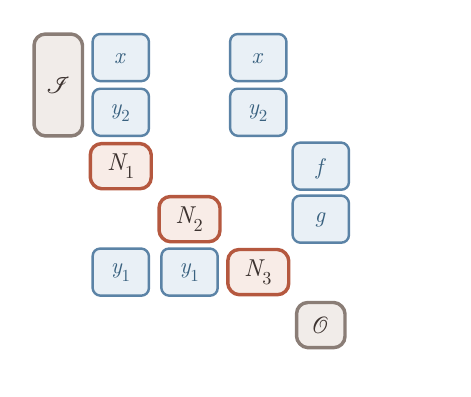}
        \caption{}\label{fig:n2_basic_left}
    \end{subfigure}
    \hfill
    \begin{subfigure}[c]{0.34\linewidth}
        \centering
        \includegraphics[width=\linewidth]{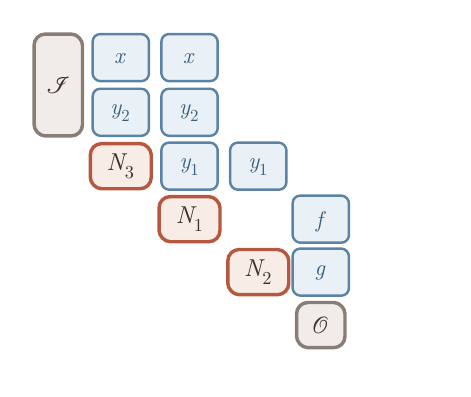}
        \caption{}\label{fig:n2_basic_right}
    \end{subfigure}
    \caption{The dependency graph given by the system of \cref{eq:sellar} in node-linking format (a) and the adjacency matrix (b).
    Inputs (not produced by any function) are denoted by the node $\mathcal{I}$, and outputs (not used as input by any function) by $\mathcal{O}$.
    The topologically sorted adjacency matrix (using as ordering top-left to bottom-right) is shown in (c), where an absence of variables under the diagonal after ordering indicates an acyclic graph.}
\end{figure}
As an example, consider the following set of pure functions (derived from the Sellar problem \cite{sellar1996response}):
\begin{alignat}{2}
    f   &= N_1(x, y_1, y_2) &&= x_1^2 + x_3 + y_1 + e^{-y_2} \label{eq:sellar}\\
    g   &= N_2(y_1)         &&= 3.16 - y_1\nonumber \\
    y_1 &= N_3(x, y_2)      &&= x_2^2 + x_3 + x_1 - 0.2 y_2 \nonumber
\end{alignat}
Since these functions are pure, only the output (first column) and inputs (second column, arguments) need to be considered for the dependency graph, shown in \cref{fig:basic_graph}, where $\mathcal{I}$ are the inputs (no function produces these values) and $\mathcal{O}$ the outputs (no function uses these values as inputs).

For a valid dependency graph, we require:
\begin{gcheck}[One producer per variable]\label{cond:one_producer}
No two nodes produce the same output variable: otherwise, the answers would be different depending on which function executed first.
\end{gcheck}
\begin{gcheck}[No self-reads]\label{cond:no_self_read}
No node takes its own output variables as inputs: such a function cannot be executed without recursion (whose termination criteria cannot be read off the graph).
\end{gcheck}

\begin{figure}[b]
    \centering
    \begin{subfigure}[c]{0.26\linewidth}
        \centering
        \includegraphics[width=\linewidth]{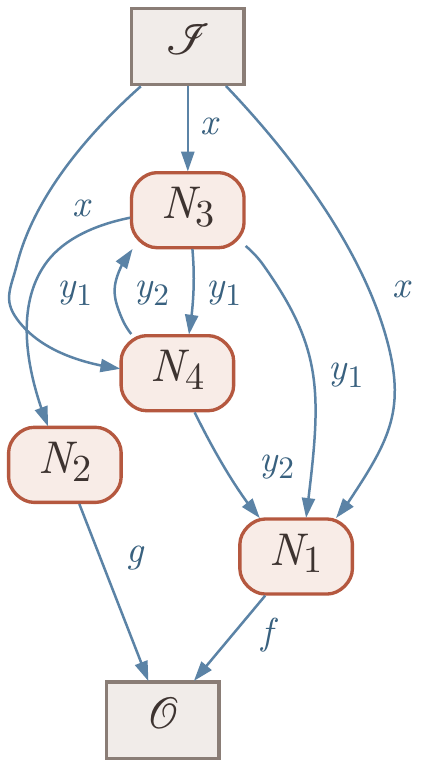}
        \caption{}\label{fig:sellar_graph}
    \end{subfigure}
    \hfill
    \begin{subfigure}[c]{0.34\linewidth}
        \centering
        \includegraphics[width=\linewidth]{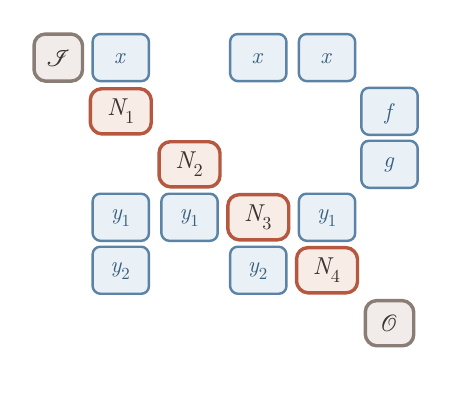}
        \caption{}\label{fig:n2_sellar_left}
    \end{subfigure}
    \hfill
    \begin{subfigure}[c]{0.34\linewidth}
        \centering
        \includegraphics[width=\linewidth]{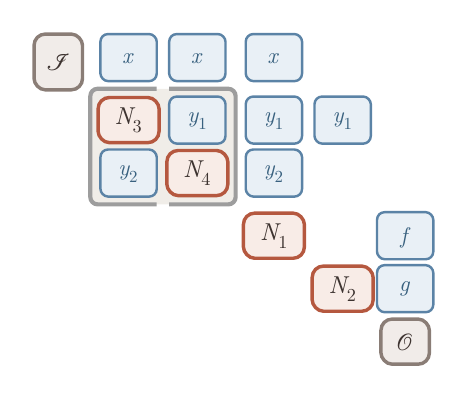}
        \caption{}\label{fig:n2_sellar_right}
    \end{subfigure}
    \caption{The cyclic dependency graph given by the system of \cref{eq:sellar} with the additional node of \cref{eq:sellar_cyclic} in node-linking format (a) and adjacency matrix format (b).
    Such cyclic graphs do not admit a feed-forward ordering, but their strongly connected components can be topologically ordered to yield an execution order in terms of components (sets of mutually dependent models or a single function), shown in (c).
    The one multi-function component $(N_3, N_4)$ is grouped in a shaded region --- no feedback is present from component to component, only within the shaded component.}
    \label{fig:sellar}
\end{figure}

A dependency graph can also be visualised in an adjacency matrix, by a matrix with each node on the diagonal\footnote{Since self-reads are excluded (\cref{cond:no_self_read}), there are no self-loops so the diagonal can be used to display the function name itself.
Furthermore, we expand the input/output nodes to fully show the variable names there too; this is not standard and as such they are differently coloured.
This is similar to the XDSM \cite{lambe2012extensions} but that focuses on the final executable graph instead of the intermediate representations used here --- it does not have a convention for the conditional nodes in \cref{sec:cycle_resolution} nor for the nesting.
Finally, the graph is always shown fully sequenced (topologically ordered strongly connected components) except in \cref{fig:n2_basic_left,fig:n2_sellar_left}} and a non-zero entry off the diagonal at $ij$ if function $i$ outputs that variable and function $j$ takes it as input, shown in \cref{fig:n2_basic_left}.

This particular graph admits a (non-unique) ordering in which each function can be evaluated: $N_3, N_1, N_2$. In general, such an ordering can be generated automatically using a topological ordering algorithm.
The adjacency matrix has an implicit ordering from top-left to bottom-right, and the topological ordering can thus be visualised in \cref{fig:n2_basic_right}.
After topological ordering, there is no variable under the diagonal, indicating no feedback and thus a valid feed-forward execution order.
However, not every graph can be topologically ordered; adding the function
\begin{alignat}{2}
    y_2 &= N_4(x, y_1)    &&= \sqrt{y_1} + x_2 + x_3 \label{eq:sellar_cyclic}
\end{alignat}
to the system of \cref{eq:sellar} yields the cyclic dependency graph shown in \cref{fig:sellar_graph}.
This graph cannot be topologically ordered directly, but a strongly connected components (SCC) algorithm yields a set of components (sets of mutually dependent nodes or a single node) that can be topologically ordered (topologically ordered strongly connected components, TOSCC) --- an execution order is then in terms of these components.
For this system, this yields
\begin{equation}
     (N_3, N_4), N_1, N_2 \label{eq:sellar_toscc}
\end{equation}
where only the first component is cyclic and has multiple nodes. 
Each cycle in the graph indicates a consistency requirement and thus the need to solve it in some way.

Summarising, if a set of pure functions and their input/outputs are defined, a dependency graph can be assembled and checked for validity (\cref{cond:one_producer,cond:no_self_read}). Its TOSCC then yields a valid execution order if acyclic, and otherwise a sequence of components, each a single function or a cycle that requires resolving (e.g. by some form of solver).

\subsection{Conditional nodes}\label{sec:cycle_resolution}
The Sellar example cannot be run directly due to its cycle (indicated by the broken component outline in \cref{fig:n2_sellar_right}). Although in this simple example it can be simply iterated until it is converged, we use this as an opportunity to introduce the \emph{conditional node} extension to the dependency graph which provides a useful unification of solvers and optimisers.

A conditional node imposes a condition on some variables and may own variables (its \emph{unknowns}) that it can vary to satisfy those conditions. These conditional nodes can (given they have unknowns and a \emph{driver}) iterate their strongly connected component in order to satisfy their conditions --- in contrast to functions which may only execute themselves. For an executable graph, the following checks are then required to pass: 

\begin{echeck}[Conditions determined]\label{cond:condition_unknowns}
Every conditional node has to have unknowns that its conditions depend on: otherwise, it cannot ensure that its conditions are satisfied when the graph is executed.
\end{echeck}

\begin{echeck}[Functionally acyclic]\label{cond:func_acyclic_2}
Only conditional nodes may receive feedback: once they are removed, the graph has to be acyclic as functions cannot make feedback consistent.
\end{echeck}

\begin{echeck}[Drivers assigned]\label{cond:driver_assigned}
Every conditional node is assigned a compatible driver (one that can satisfy the condition: an optimiser driver on an optimisation-type condition), which varies its unknowns until its conditions are satisfied.
\end{echeck}

\begin{figure}[b]
    \centering
    \includegraphics[width=\linewidth]{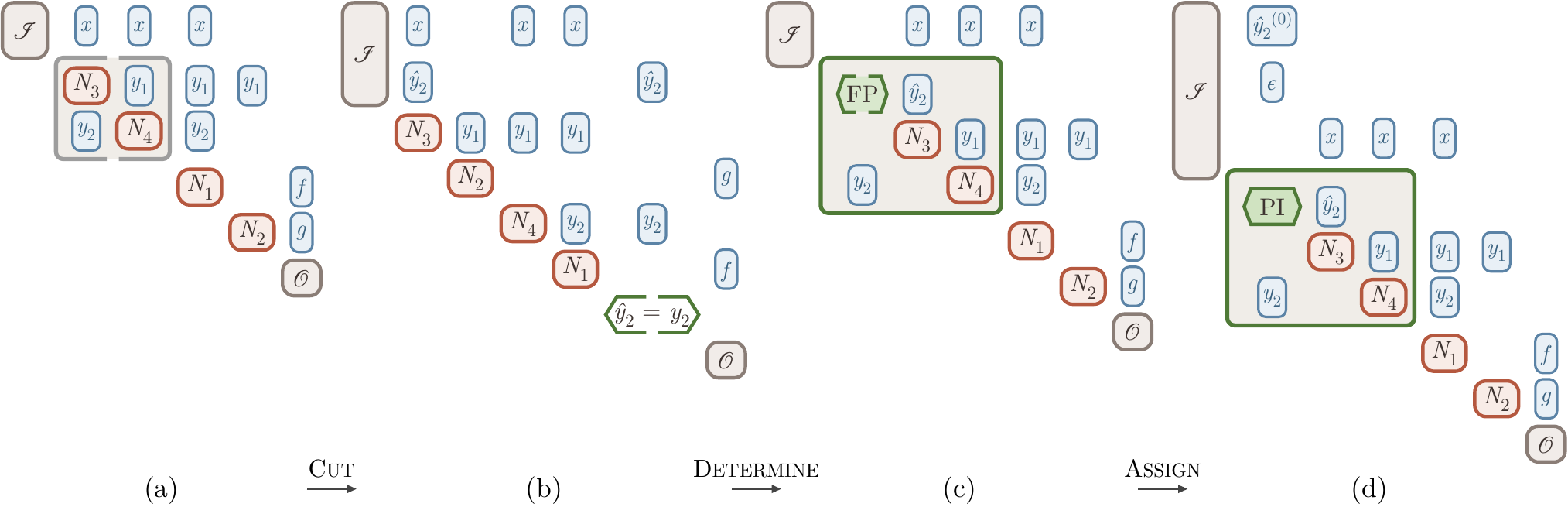}
    \caption{The Sellar example of \cref{fig:sellar}: (a) initial, (b) after copying the output $y_2$ of $N_4$ into $\hat{y}_2$ for $N_3$ to read, with the condition $\hat{y}_2 = y_2$ stated, (c) after the conditional node determines $\hat{y}_2$, making it a fixed-point (FP) node, (d) after assigning FP the compatible Picard iteration (PI) driver. 
    This Picard iteration requires an initial value $\hat{y}_2^{(0)}$ and a tolerance $\epsilon$ to execute.
    An unresolved function cycle component is indicated with a broken outline, an undetermined condition (without unknowns) is indicated with a broken hexagonal outline (coloured by the type of condition), a condition with unknowns is indicated with a filled hexagon with broken outline, and an assigned driver is a filled hexagon.
    The last graph is fully executable.}
    \label{fig:sellar_cuts}
\end{figure}

In the Sellar example, a functional node has feedback in its SCC, and this has to be converted to a graph with a conditional node obtaining that feedback and a driver iterating it. The chain of operations doing so is shown in \cref{fig:sellar_cuts}.

First, a copy of $y_2$ ($\hat{y}_2$) is made (only read by the function in the cycle that reads it, $N_3$) and a conditional node in the graph is inserted that requires these to be equal (together, a \op{Cut}). This does not change the solution of the graph --- replacing $\hat{y}_2 \to y_2$ yields the original graph again. \Cref{cond:condition_unknowns,cond:driver_assigned} require it to be given unknowns and a driver to make the graph executable. The \op{Determine} operation sets the unknowns of the conditional node to $\hat{y}_2$. The \op{Assign} operation then assigns the Picard iteration to the conditional node. 

The graph of \cref{fig:sellar_cuts}(d) satisfies \cref{cond:condition_unknowns,cond:func_acyclic_2,cond:driver_assigned}: the Picard iteration node has unknowns, it is the only node receiving feedback (under the diagonal, in contrast to \cref{fig:n2_sellar_right} where the functional node $N_3$ receives feedback), and it has been assigned a driver.

The assigned driver may also use external inputs that are not part of its conditions such as an initial guess $\hat{y}^{(0)}_{2}$ or a tolerance $\epsilon$, and output other quantities such as iteration history. These remain part of the dependency graph and can be wired to other functions: an initial guess may come from a low-fidelity variant and its tolerance may be scaled by the other inputs. However, they cannot be part of what a driver iterates: otherwise they \textit{are} just conditions or unknowns since a driver by definition iterates only those.

\begin{figure}[h]
    \begin{subfigure}[t]{0.248\linewidth}
        \centering
        \includegraphics[width=0.977\linewidth]{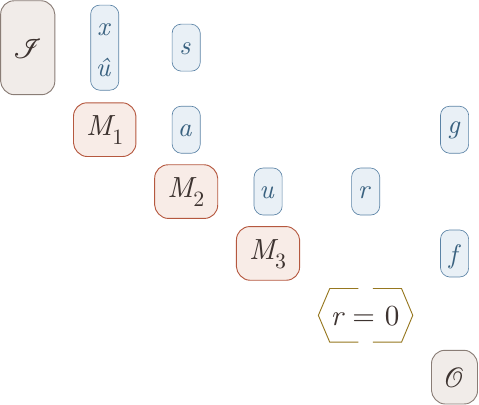}\\[2pt]
        {$\downarrow$\;\op{Determine}\ $s$}\\[2pt]
        \includegraphics[width=1.000\linewidth]{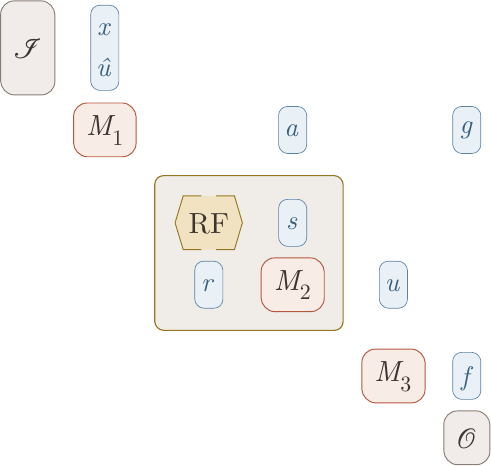}
        \caption{}\label{fig:determine_rf}
    \end{subfigure}
    \hfill
    \begin{subfigure}[t]{0.247\linewidth}
        \centering
        \includegraphics[width=0.994\linewidth]{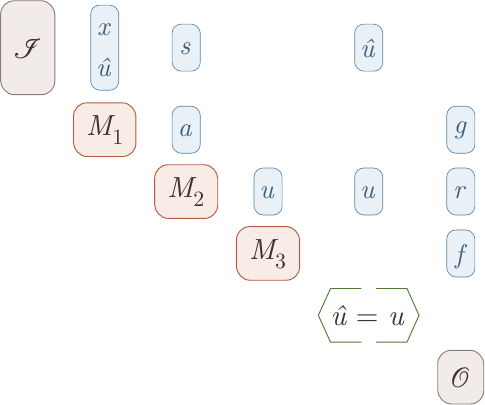}\\[2pt]
        {$\downarrow$\;\op{Determine}\ $\hat{u}$}\\[2pt]
        \includegraphics[width=1.000\linewidth]{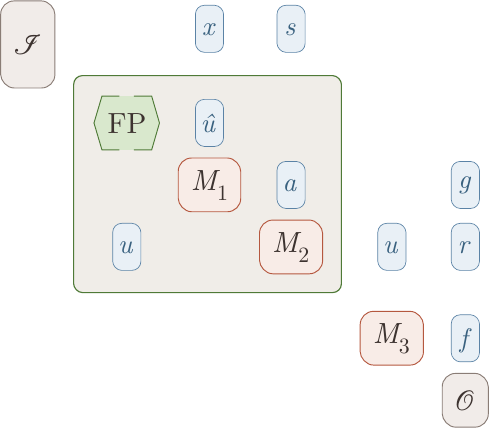}
        \caption{}\label{fig:determine_fp}
    \end{subfigure}
    \hfill
    \begin{subfigure}[t]{0.296\linewidth}
        \centering
        \includegraphics[width=0.911\linewidth]{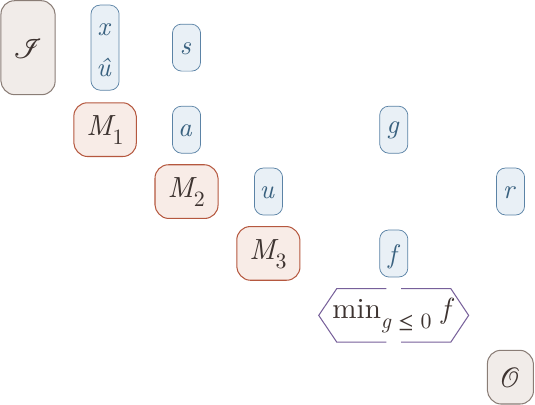}\\[2pt]
        {$\downarrow$\;\op{Determine}\ $x$}\\[2pt]
        \includegraphics[width=1.000\linewidth]{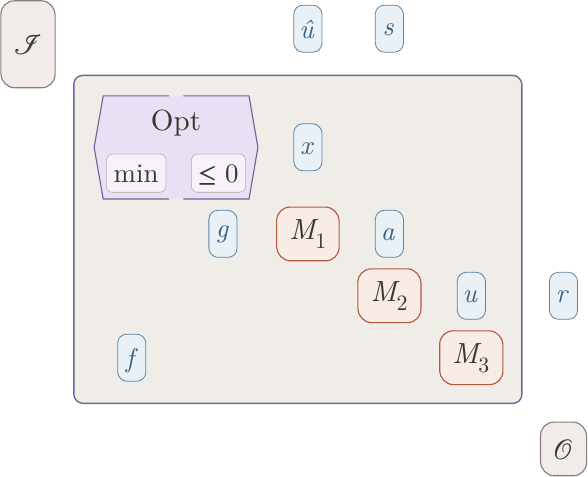}
        \caption{}\label{fig:determine_opt}
    \end{subfigure}    
    \caption{Three different types of conditions each determining a different variable in the same dependency graph: (a) a root find, (b) a fixed point, and (c) an optimisation.
    The size of the induced cycle depends on the determined variable.}    
    \label{fig:problem_types}
\end{figure}

Conditional nodes are general: for example, in \cref{fig:problem_types} three condition types are shown in the graph before and after determining. The minimum set of models that they have to iterate immediately follows from the TOSCC. Optimisation, fixed point, and root find are, from the graph's perspective, all a condition determining a variable (the colouring is for visualisation purposes only) and the executability checks hold in the same way whatever the type of conditions are in the graph. Furthermore, the noted graph operations in \cref{fig:sellar_cuts} apply to all of them (but the type of conditional node determines which drivers can be assigned to it), as well as the nesting and combining of such nodes discussed in the next section.

\subsection{Nesting and combining}\label{sec:nest_combine}
\begin{figure}[h]
    \centering
    
    \includegraphics[width=0.719\linewidth]{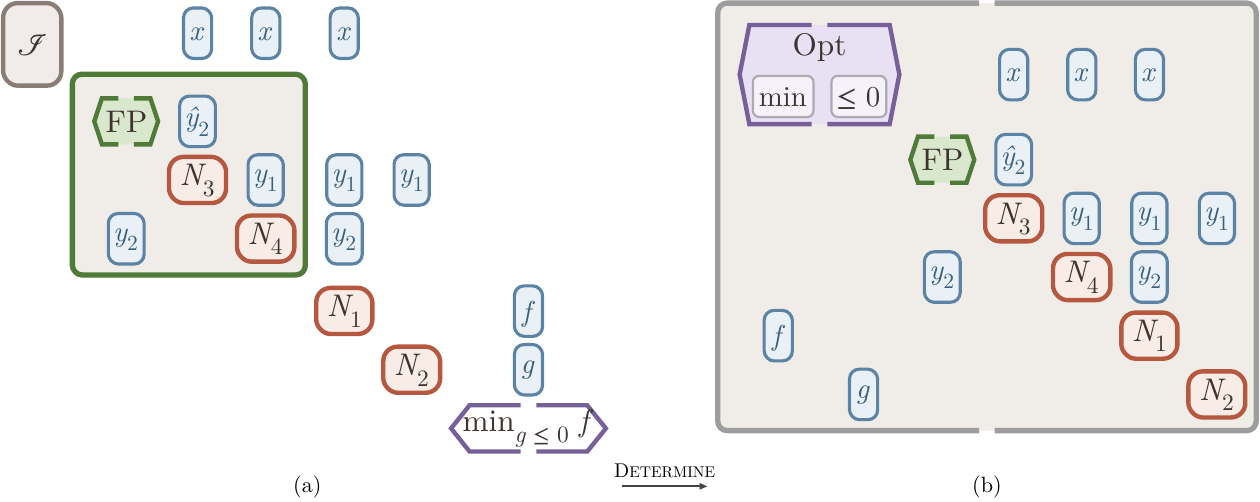}    
    \caption{The Sellar problem in \cref{fig:sellar_cuts}(c), with an added optimality condition (a), which then determines the design variable $x$, resulting in one strongly connected component with two conditional node types (b).}\label{fig:determine}
\end{figure}

Multiple conditional nodes can be present in the same strongly connected component: for example, adding an optimality condition determining $x$ to the Sellar problem of \cref{fig:sellar_cuts} yields one component containing both the fixed-point node and the optimiser node, as shown in \cref{fig:determine}. 

\begin{figure}[h]
    \centering
    \includegraphics[width=\linewidth]{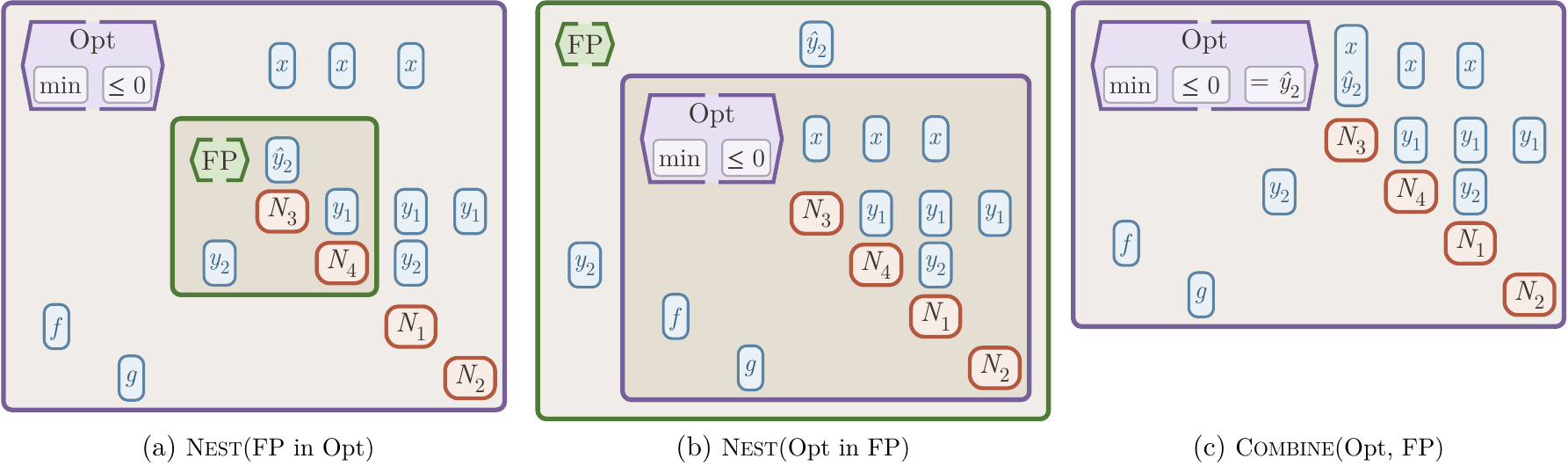}
    \caption{Three ways to make \cref{fig:determine}(b) executable: in (a) the fixed-point conditional node is nested such that the optimiser executes it in each iteration, in (b) the fixed point is outer and executes the optimiser per iteration, and in (c) the two conditional nodes are combined. (a) and (c) have the same solution sets; (b) a different one.}
    \label{fig:nesting}
\end{figure}

The TOSCC defines no order within an SCC and as such the graph itself does not define whether the optimiser iterates the fixed-point node or vice versa (in \cref{fig:determine} by convention the optimiser is outer, \cref{fig:nesting}(a) and (b) show the two options). This in general leads to different solution sets and as such we have to force an explicit decision on which ordering is meant. 

This is in the form of a \emph{nesting edge} added to the dependency graph between two conditional nodes. It states the first node is nested within the second, with the first conditional node being fully satisfied for each iteration of the second. For a valid dependency graph with nesting edges, it has to additionally satisfy:

\begin{gcheck}[No nesting loops]\label{cond:no_nesting_loops}
No conditional node is nested within itself, not even through other conditional nodes (no $A\to B\to A$).
\end{gcheck}

and for that graph to be executable, it has to hold that:

\begin{echeck}[Valid nesting]\label{cond:valid_nesting}
Every component with multiple conditional nodes has exactly one outermost conditional node, the same holding within that component's subgraph after removing the outermost conditional node. Nesting edges also only connect conditional nodes sharing a component, as nesting is meaningless otherwise.
\end{echeck}

In \cref{fig:nesting} the three options for making \cref{fig:determine}(b) executable are shown: either the fixed point is \op{Nest}ed within the optimiser, or vice versa, or the two conditional nodes are \op{Combine}d together, adding their conditions. This may change the combined node's condition type: here, the fixed point becomes a constraint on the optimiser. (a) and (c) will yield the same solution set while (b) will in general not.

In nested figures, the ordering is still from top-left to bottom-right. If an iterated component (indicated by a solid line outline) is encountered, that is first fully converged before moving on. As an example, in \cref{fig:nesting}(a), the purple box indicates it will be iterated. The top-left node is the optimiser, then comes the green solid box: the fixed-point node is top-left within that, so it is first converged fully (by iterating $N_3$ and $N_4$) before moving to the next node $N_1$ and then $N_2$.

\subsection{Full workflow}\label{sec:workflow}
A collection of models and nesting edges defines a dependency graph, which is considered valid if \cref{cond:one_producer,cond:no_self_read,cond:no_nesting_loops} pass. If these do not pass, the set of models has to be changed or the problematic nesting edges have to be removed. Then, for a fully executable graph, \cref{cond:condition_unknowns,cond:func_acyclic_2,cond:driver_assigned,cond:valid_nesting} have to be satisfied: a valid graph can be converted to an executable graph by using the graph operations \op{Cut} (\cref{cond:func_acyclic_2}), \op{Determine} (\cref{cond:condition_unknowns}), \op{Nest} or \op{Combine} (\cref{cond:valid_nesting}), and \op{Assign} (\cref{cond:driver_assigned}).

\begin{figure*}[h]
    \centering
    \def\wffont{\fontsize{7}{8.2}\selectfont}%
\def\wfnote{\fontsize{6}{7}\selectfont}%
\definecolor{c_wf_ink}{RGB}{61,51,48}%
\definecolor{c_wf_note}{RGB}{125,113,108}%
\definecolor{c_wf_model_f}{HTML}{f8ece7}\definecolor{c_wf_model_s}{HTML}{b5583f}%
\definecolor{c_wf_data_f}{HTML}{e9f0f6}\definecolor{c_wf_data_s}{HTML}{5b83a6}%
\definecolor{c_wf_group_f}{HTML}{eae6df}\definecolor{c_wf_group_s}{HTML}{8b8173}%
\definecolor{c_wf_prob_f}{HTML}{d9d9d9}\definecolor{c_wf_prob_s}{HTML}{5e5e5e}%
\definecolor{c_wf_prob_det}{HTML}{e6e6e6}\definecolor{c_wf_prob_drv}{HTML}{e1e1e1}%
\definecolor{c_wf_check}{HTML}{96771f}%
\begin{tikzpicture}[
    x=1cm, y=1cm, font=\wffont, >={Stealth[length=3.5pt, width=3pt]},
    every node/.style={align=center},
    box/.style={draw=c_wf_ink, line width=0.55pt, fill=white, rounded corners=2.5pt,
                inner xsep=6pt, inner ysep=3.5pt, minimum height=7mm, minimum width=1.7cm},
    model/.style={box, draw=c_wf_model_s, fill=c_wf_model_f},
    graph/.style={box, draw=c_wf_group_s, fill=c_wf_group_f},
    op/.style={box, draw=c_wf_data_s, fill=c_wf_data_f, inner xsep=3pt},
    hex/.style={draw, line width=0.55pt, signal, signal from=nowhere,
                signal to=east and west, signal pointer angle=115,
                inner xsep=7pt, inner ysep=3.5pt, minimum height=7mm, minimum width=1.7cm},
    stated/.style={hex, draw=none, fill=white},
    determined/.style={hex, draw=none, fill=c_wf_prob_det},
    driven/.style={hex, draw=c_wf_prob_s, fill=c_wf_prob_drv},
    check/.style={draw=c_wf_check, line width=0.55pt, fill=white, diamond, aspect=0.8,
                  inner sep=0.5pt},
    note/.style={font=\wfnote, text=c_wf_note, align=center, inner sep=1pt},
    flow/.style={->, draw=c_wf_ink, line width=0.6pt},
    branch/.style={->, draw=c_wf_ink, line width=0.6pt, rounded corners=4pt},
    lab/.style={font=\wfnote, text=c_wf_ink, inner sep=1.5pt},
]
\providecommand{\op}[1]{\textsc{#1}}%
\let\opname\op%
\def\see#1{{\wfnote\textcolor{c_wf_note}{#1}}}%
\def\cutedge#1{%
  \draw[c_wf_prob_s, line width=0.55pt] ($(#1.north)+(-1.75pt,0)$) -- (#1.north west) -- (#1.west) -- (#1.south west) -- ($(#1.south)+(-1.75pt,0)$);%
  \draw[c_wf_prob_s, line width=0.55pt] ($(#1.north)+(1.75pt,0)$) -- (#1.north east) -- (#1.east) -- (#1.south east) -- ($(#1.south)+(1.75pt,0)$);}%
\node[model]      (fn)   at (-0.3, 1.65)  {functions};
\node[stated]     (cond) at (-0.3, 0.6)  {conditional\\nodes\\\see{no unknowns}};
\node[determined] (prob) at (-0.3,-0.55)  {conditional\\nodes\\\see{with unknowns}};
\cutedge{cond}\cutedge{prob}%
\node[driven]     (drv)  at (-0.3,-1.7)  {conditional\\nodes\\\see{with driver}};
\node[note] (dtitle) at (-0.3, 2.2) {declared at any stage};
\node[draw=c_wf_note, dashed, line width=0.45pt, rounded corners=3pt, inner sep=5pt,
      fit=(dtitle) (fn) (cond) (prob) (drv)] (decl) {};
\def\cond#1{\see{\ref{#1}}}%
\def\condrange#1#2{\see{\ref{#1}--\ref{#2}}}%
\node[check, right=0.25cm of decl] (chk0) {valid\\graph?\\\condrange{cond:one_producer}{cond:no_nesting_loops}};
\node[graph, right=0.3cm of chk0, minimum width=0pt, inner xsep=4pt] (asm)
    {graph\\TOSCC\\\see{fig.~\ref{fig:sellar}}};
\node[check, right=0.3cm of asm]   (chk1) {functionally\\acyclic?\\\cond{cond:func_acyclic_2}};
\node[check, right=0.4cm of chk1]  (chk2) {conditions\\determined?\\\cond{cond:condition_unknowns}};
\node[check, right=0.4cm of chk2]  (chk3) {valid\\nesting?\\\cond{cond:valid_nesting}};
\node[check, right=0.4cm of chk3]  (chk4) {drivers\\assigned?\\\cond{cond:driver_assigned}};
\node[graph, right=0.3cm of chk4]  (exe)  {execute};
\node[op] (cut)  at (chk1 |- 0, 2.3) {\opname{Cut}\\\see{fig.~\ref{fig:sellar_cuts}(b)}};
\node[op] (det)  at (chk2 |- 0, 2.3) {\opname{Determine}\\\see{fig.~\ref{fig:determine}}};
\node[op] (nest) at (chk3 |- 0, 2.3) {\opname{Nest}\,$\cdot$\,\opname{Combine}\\\see{fig.~\ref{fig:nesting}}};
\node[op] (asg)  at (chk4 |- 0, 2.3) {\opname{Assign}\\\see{fig.~\ref{fig:sellar_cuts}(d)}};
\draw[flow] (decl.east |- chk0) -- (chk0.west);
\draw[flow] (chk0) -- node[lab, above, pos=0.5] {yes} (asm);
\draw[flow] (asm)  -- (chk1);
\draw[flow] (chk1) -- node[lab, above, pos=0.5] {yes} (chk2);
\draw[flow] (chk2) -- node[lab, above, pos=0.5] {yes} (chk3);
\draw[flow] (chk3) -- node[lab, above, pos=0.5] {yes} (chk4);
\draw[flow] (chk4) -- node[lab, above, pos=0.5] {yes} (exe);
\foreach \c/\o in {chk1/cut, chk2/det, chk3/nest, chk4/asg}
    \draw[branch] (\c.north) -- node[lab, right, pos=0.45] {no} (\o.south);
\coordinate (err) at ($(decl.south) + (0, -0.4)$);
\draw[branch, dashed, draw=c_wf_check] (chk0.south) -- node[lab, right, pos=0.3] {no} (chk0.south |- err)
    -- (err) -- (decl.south);
\coordinate (ret) at ($(asg.north) + (0, 0.45)$);
\foreach \o in {cut, det, nest} \draw[branch, -] (\o.north) -- (\o.north |- ret);
\draw[branch] (asg.north) -- (ret) -- (decl.north |- ret) -- (decl.north);
\end{tikzpicture}%
    \caption{Schematic workflow of obtaining an executable graph from a set of functions and conditional nodes --- all checks need to pass before executing.
    First, the dependency graph implied by this set is checked for validity by \cref{cond:one_producer,cond:no_self_read,cond:no_nesting_loops} (if it is not valid, the declarations need to be amended), and its TOSCC is computed.
    Function cycles (functions depending on each other's output) have to be \op{Cut} (which adds a conditional node, \cref{cond:func_acyclic_2}), conditional nodes without unknowns that they depend on have to be \op{Determine}d (\cref{cond:condition_unknowns}), conditional nodes sharing a component have to be correctly nested or combined (\cref{cond:valid_nesting}), and finally conditional nodes without a driver need to be assigned one (\cref{cond:driver_assigned}).
    Each operation amends the declarations, after which the checks are repeated.
    If all checks pass, the graph can be executed.
    The order does not have to be the one drawn: each conditional node is independent and could be assigned a driver before a function cycle is cut.
    Furthermore, an acyclic set of functions (without conditional nodes) would immediately pass all checks and can be executed after TOSCC.}
    \label{fig:workflow}
\end{figure*}
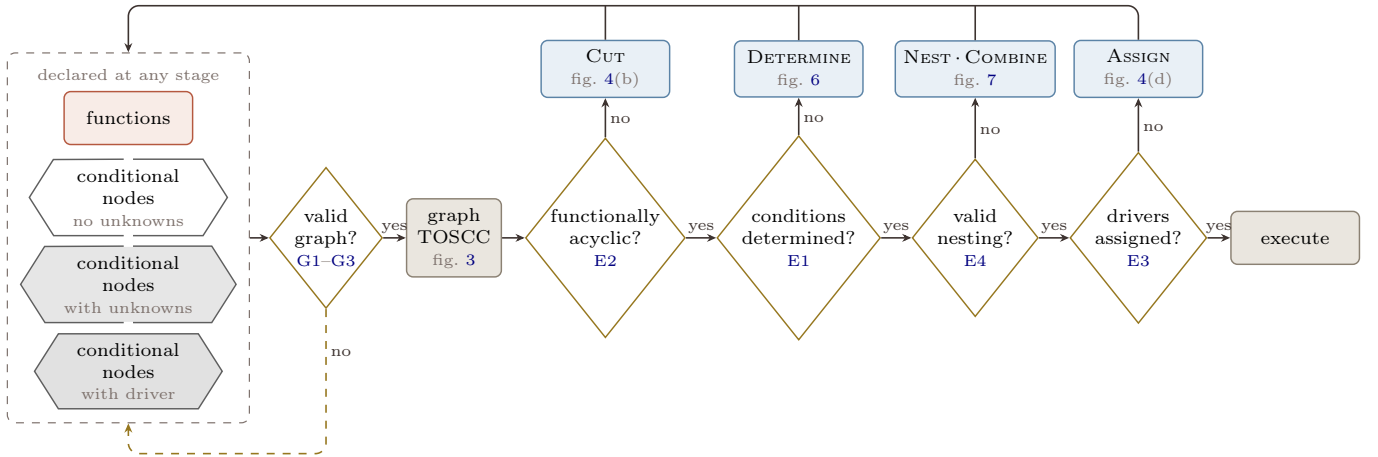

The full workflow for turning a set of models and additional conditional nodes (such as optimality conditions or consistency constraints) into an executable dependency graph is shown in \cref{fig:workflow}.
This does not have to be as linear as shown here: the graph contains functions and conditional nodes with or without unknowns and drivers at the same time; this is merely an order that is commonly encountered. It is possible to determine new design variables after obtaining an executable graph or start with already assigned drivers for e.g. implicit models.
Finally, note that a set of acyclic functions with non-overlapping outputs would pass all checks immediately and is simply ordered and executed.

The workflow of \cref{fig:workflow} can be used for arbitrary models and can create a highly customised MDAO architecture (addressing \cref{req:selection,req:flexible}). However, a sequence of manually chosen graph operations applies to one set of models and has to be redone whenever the design problem changes; \cref{req:coordination} would require at least an automated starting point for an MDAO architecture. This is the topic of the next section.

\subsection{MDAO architectures as graph operations}\label{sec:architectures}
Such starting points can be provided by the standard architectures in the MDAO literature~\cite{martins2013multidisciplinary}. Monolithic architectures (with a single top-level optimisation problem) can be implemented in terms of rule-based sequences of the graph operations in \cref{fig:workflow}, allowing for immediately and automatically obtaining an executable graph for any given design optimisation problem. After this, the graph operations can be used to modify this basic architecture to increase efficiency or convergence properties. Distributed architectures \cite{tosserams_classification_2009} require inserting additional functional nodes and optimisers; monolithic architectures do not.

\begin{figure}[t]
    \centering    
    \includegraphics[width=0.740\linewidth]{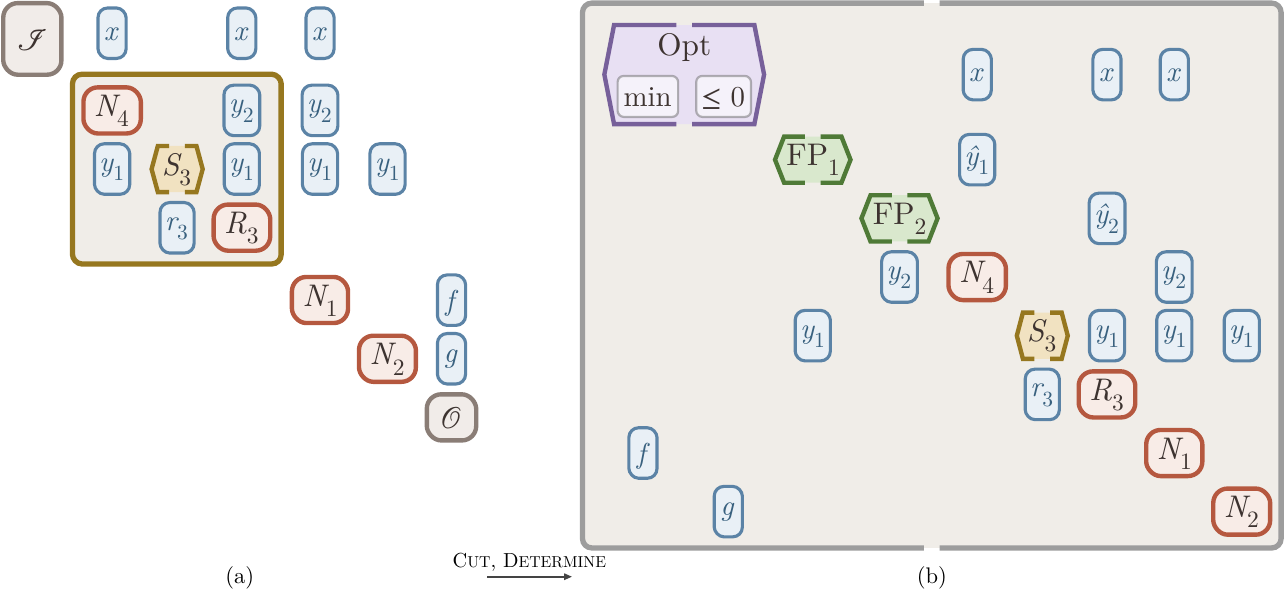}    
    \caption{The Sellar problem with an additional implicit component: $N_3\to S_3, R_3$ (a).
    The coupling variables $y_1$ and $y_2$ (produced in that SCC and read by a functional node) are cut and determined and the optimiser is reinserted, yielding four distinct conditional nodes in the same component (b).} 
    \label{fig:architectures_statement}
\end{figure}

The recipe for a monolithic architecture will be illustrated using the extended Sellar problem, where the $N_3$ explicit function is replaced by an implicit function (a residual $r_3$ and a root-find condition determining $y_1$ from that residual) and an optimiser minimises $f$ with $g\leq0$ by varying $x$.
This is shown without the optimiser in \cref{fig:architectures_statement}(a).

\begin{figure}[t]
    \centering    
    \includegraphics[width=\linewidth]{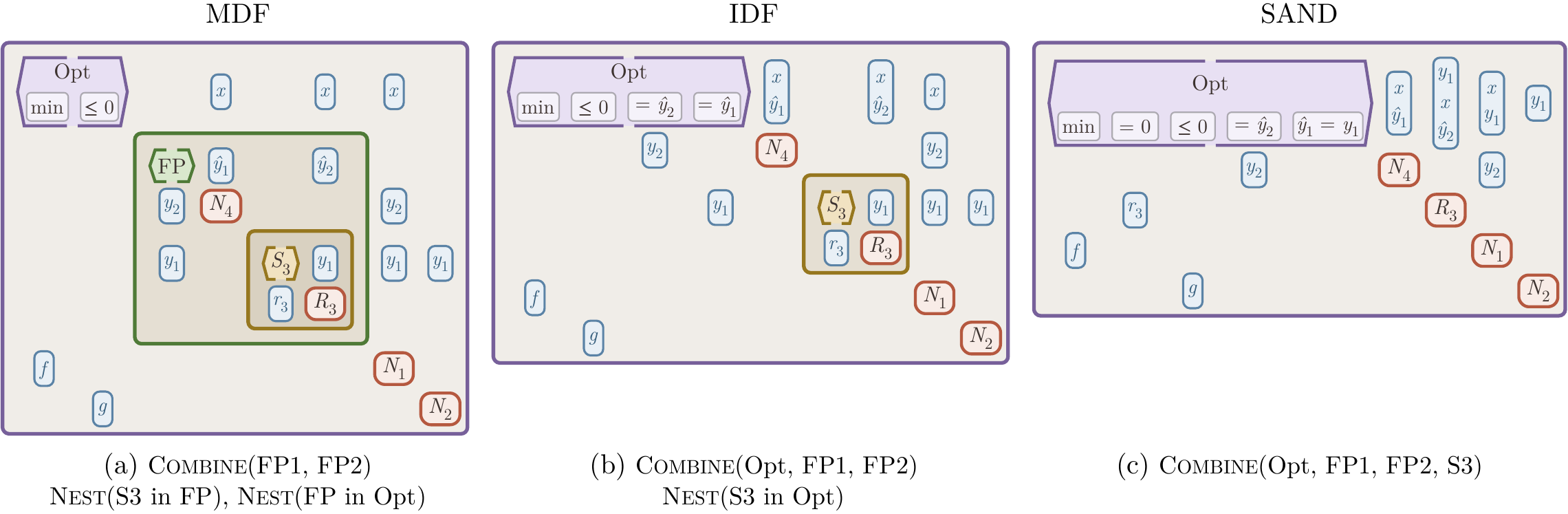}    
    \caption{Three MDAO architectures for the Sellar problem with the combining and nesting operations producing them from \cref{fig:architectures_statement}(b): (a) a multidisciplinary-feasible architecture which resolves all coupling in an inner multidisciplinary analysis, (b) an individual-discipline-feasible architecture which resolves coupling globally but the model solve locally, and (c) a simultaneous-analysis-and-design architecture which resolves both coupling and model solves globally in the same optimiser.
    In (c), the $\hat{y}_1 = y_1$ condition becomes trivial and can be removed (the optimiser owns both of them). }
    \label{fig:architectures}
\end{figure}

\begin{enumerate}
    \item Remove the optimiser and compute the TOSCC to find the couplings within each optimiser iteration.
    \item \op{Cut} and \op{Determine} the functional feedback to satisfy \cref{cond:func_acyclic_2}. Possible options are to \op{Cut} all coupling variables in the component except to the solvers ($S_3$ in \cref{fig:architectures_statement}) --- Jacobi --- or use Gauss--Seidel (which requires ordering) to cut a smaller set. See \cref{fig:architectures_statement}(b) for the Jacobi version that cuts both $y_1$ and $y_2$.
    \item Reinsert the optimiser node, \op{Combine} and \op{Nest} the original solvers (root finders, $S_3$ in \cref{fig:architectures_statement}), the obtained fixed-point conditional nodes, and the optimiser node in some way; see \cref{fig:architectures} for three examples.
    \item \op{Assign} all conditional nodes compatible drivers from a set of defaults (e.g. Newton for root finders, Picard for fixed points).
\end{enumerate}

The multidisciplinary-feasible architecture (MDF) is shown in \cref{fig:architectures}(a): the two fixed-point iterations are combined, the original solver $S_3$ is nested in that fixed-point iteration, which is nested in the optimiser. The individual-discipline-feasible architecture (IDF) in \cref{fig:architectures}(b) combines the fixed-point iterations with the optimiser while nesting $S_3$ in the optimiser. Finally, the simultaneous-analysis-and-design architecture (SAND) in \cref{fig:architectures}(c) combines all conditional nodes in the top-level optimiser.

These monolithic architectures differ only in the nesting level at which the implicit models and the coupling between models (i.e. cycles in the graph) are resolved: locally with a solver, medium-level with a multidisciplinary analysis (MDA), or globally by the optimiser.
The \op{Nest} and \op{Combine} operations are how these choices are made.

Although this example is simple, the same strategy can be used for an arbitrarily complex set of models, as will be done in the PROCESS case study (\cref{sec:process}), where exactly these recipes for MDF and SAND are used for a set of over 140 interconnected models. 

Finally, these classifications are not exhaustive: an MDAO architecture can arbitrarily be built from locally and globally resolved couplings and models (e.g. the MDF architecture could as well root-find the entire Sellar system at once by combining the FP and the $S_3$), which these graph operations support --- in other words, these operations make it possible to mix and match elements from any architecture (e.g. almost all couplings globally except some difficult local solves), addressing \cref{req:coordination,req:flexible}.

\subsection{Summary}
In engineering design and optimisation, the starting point is often a set of explicit models, implicit models (defined as a solver and a residual computation) and conditional nodes (which include both consistency constraints and optimality conditions). 

If these are all built from pure functions with explicit inputs and outputs, they define a dependency graph of which the validity can be checked using \cref{cond:one_producer,cond:no_self_read,cond:no_nesting_loops}. This graph yields a valid execution order in terms of components: either a single function (in which case it can be executed), or a component with multiple functions and conditional nodes (cyclic), in which case it has to be resolved. 

The graph operations (\op{Cut}, \op{Determine}, \op{Nest}, \op{Combine}, \op{Assign}) can be used to manipulate the graph until the executability \cref{cond:condition_unknowns,cond:func_acyclic_2,cond:driver_assigned,cond:valid_nesting} pass. Finally, as shown in \cref{sec:architectures}, sequences of graph operations can be used to automatically assign an MDAO architecture; the graph operations can then be used to modify these standard options to better suit the problem at hand.

This graph-based theory addresses \cref{req:selection,req:coordination,req:flexible}: arbitrary design problems (e.g. different design variables are supported by changing the unknowns of the optimiser) can be used, an automated MDAO architecture can be defined for any design problem, and this can be flexibly modified using graph operations. \Cref{req:gradients,req:overhead} are addressed in the following section, which discusses the implementation of the graph-based theory in \cottax{}.
\clearpage

\section{Implementation}\label{sec:imp}
The graph-based theory and extension of the dependency graph of \cref{sec:methods} are implemented in the Python package \cottax{}.
It is designed to be extensible: it should support new types of conditional nodes (e.g. least-squares / derivative conditions), data structures (e.g. PyTrees \cite{jax2018github} or something entirely different), and use-cases (\textit{consumers}, e.g. exporting or running the graph, which may depend on each other). 

\begin{figure}[h]
    \centering
    \includegraphics[width=0.552\linewidth]{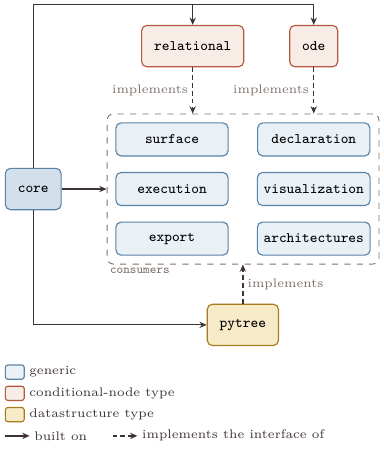}
    \caption{Software architecture of \cottax{}.
    Blue indicates generic in both conditional node types and data structure types.
    Red indicates a conditional node type and yellow a data structure type (they are generic in the other).
    These implement the interface of the consumers: the exporting (\cref{sec:exporting}), the declaration and surface interface (\cref{sec:interfaces}), the visualisation (\cref{sec:visualisation}), the architectures of \cref{sec:architectures} (\cref{sec:architectures_cottax}), and the execution (\cref{sec:execution}).
    }
    \label{fig:layers}
\end{figure}

The \texttt{core} module of the package is then generic: the mathematical graph checks and operations on it are written entirely in plain Python.
The current \textit{consumers} are exporting (\cref{sec:exporting}), declaring (and manipulating and running) the graph with a convenient interface (\cref{sec:interfaces}), visualising (\cref{sec:visualisation}), using the architectures of \cref{sec:architectures} (\cref{sec:architectures_cottax}), and executing the graph using JAX (\cref{sec:execution}).
Currently, the implemented conditional node types are relational and (experimental) ordinary differential equations (which can be combined in the same graph). PyTrees (a nested hierarchical Python structure) are the implemented data structure types.
The software architecture is shown in \cref{fig:layers}.
Each use-case declares what it requires from a node or variable place (\cref{app:places}) type to work (e.g. the JSON export declares that it needs a way to convert a place to a string).
The node or variable place extension registers its interface implementation if desired (dependency inversion).
A final assembly file can define shorthand for the consumers (with e.g. some defaults already filled in and a place and conditional node types selected).
Each of the consumers and types will be briefly described below.

The \cottax{} package is available open-source with an MIT licence at \url{https://github.com/timobogaarts/cottax}.

\subsection{Relational}\label{sec:relational}
Relational conditional node types unify common types of MDAO solvers in one node type that can be combined and classified.

A relation is defined as comparing two quantities in the graph (e.g. $a = b$ or $a \leq b$, where $b$ is either another variable or a number) and the conditional node type combines these relations into one node type. 

Conditional nodes can then be classified by their relations and objective once the unknowns are chosen as in \cref{tab:conditions}.
These conditions can be simply combined by adding the set of relations, unknowns, and objective(s), which may change their classification and thus their compatible drivers.

\begin{table}[h]
    \caption{Relational conditional nodes classified as a problem type by the relations and objective in their condition. $x$ is the unknown (one or more); $a$, $b$ and $f$ are variables that depend on $x$ through the nodes between them (note these can all be written as $c(x) = 0$ on the right-hand side without loss of generality). $\checkmark$: present, $(\checkmark)$: may be present.
    Fixed point and root find only differ in that the unknown $x$ is present directly in the fixed-point problem.}
    \label{tab:conditions}
    \centering
    \setlength{\tabcolsep}{9pt}
    \begin{tabular}{@{}lccccl@{}}
        \toprule
                                   & \multicolumn{3}{c}{Relations}                  & Objective    &                                     \\
        \cmidrule(lr){2-4} \cmidrule(lr){5-5}
        Problem                    & $x = b(x)$   & $a(x) = b(x)$  & $a(x) \leq b(x)$ & $\min f(x)$ & Drivers (examples)                  \\
        \midrule
        Fixed point                & $\checkmark$ &                &                &              & Picard, Anderson\footnote{Jacobi and Gauss--Seidel are not drivers but rather a different dependency graph: they differ in the readers of the copy $\hat{y}_2$: for Jacobi all, for Gauss--Seidel only the minimal set for a particular ordering.}                   \\
        Root find                  &              & $\checkmark$   &                &              & Newton, Broyden                     \\
        Feasibility                &              & $(\checkmark)$ & $\checkmark$   &              & constrained optimiser, no objective \\
        Unconstrained optimisation &              &                &                & $\checkmark$ & BFGS, Nelder--Mead                  \\
        Constrained optimisation   &              & $(\checkmark)$ & $(\checkmark)$ & $\checkmark$ & SLSQP, VMCON, IPOPT                 \\
        \bottomrule
    \end{tabular}
\end{table}
\subsection{Ordinary differential equations}\label{sec:ode}
Ordinary differential equations can be computed algebraically (using a residual-like formulation) using the relational conditional node types through collocation (see e.g. Dymos \cite{falck2021dymos}), but this module implements a rate conditional node type for initial value problem integration.

This conditional node satisfies its condition when it has integrated this rate to the specified final time (by iterating its subgraph).
The trajectory can be optionally returned.
Solvers such as Diffrax~\cite{kidger2021on} can then be used.
These conditional node types can be used within the same dependency graph (e.g. a root find can be nested in a rate cycle).

\subsection{PyTree}\label{sec:pytree}
A PyTree \cite{jax2018github} is an arbitrarily nested hierarchical Python data structure.
Each segment is then an index in a data structure (e.g. a \texttt{DictKey} for a dictionary key, a \texttt{GetAttrKey} for attribute access, and a \texttt{SequenceKey} for list and tuple indexing).
Both node labels and variable places are then in terms of the PyTree keys, enabling the use of any Python structure built off of the primitive indexes (and others can be added) for both the function definition tree and the data structure. 

\subsection{Exporting}\label{sec:exporting}
The graph is in all cases a set of labelled nodes with input and output names and the nesting edges, possibly with some implementation object, and can be readily exported.
Currently, two exports are available: a standard pickle, and a JSON that serialises the entire graph to strings.

The pickle and JSON are designed to be lossless: a graph can be exported and then reloaded (to separate construction of the graph from execution or visualisation). For executable graphs, this requires the implementations to be constructible from a reference (i.e. forbidding closures, unnamed lambdas, and local functions). Non-Python functions (e.g. references to functions in different languages or repository locations) can be used and exported as long as they are serialisable; this is useful as a starting point for exporting to different execution or inspection backends in the future.

\subsection{Declaration and surface}\label{sec:interfaces}
The \texttt{core} module is by construction completely unambiguous: all types are fully specified and there are on purpose no implicit conversions (e.g. ``this string actually means a dictionary key''). This makes it relatively light and maintainable: there are no such conveniences to maintain or extend in the \texttt{core} module. 

However, declaring, manipulating, and executing the graph in many cases requires referring to nodes and variables.
Furthermore, while declaring the graph, one should be able to give explicit values (a tolerance, initial values, or optimisation bounds) or additional information (e.g. units, runtime, shapes) that have no place in the base graph.
In principle, the \texttt{core} module supports these use-cases, but it can be inconvenient to always use the \cottax{}-native types or keep a parallel structure with all additional information.
To keep the \texttt{core} module light, extensible, and fast, these conveniences are instead in the \texttt{surface} and \texttt{declaration} consumers.
They make this significantly easier by providing a way to implicitly convert arguments (e.g. a ``function means its outputs'', or ``a string refers to getting that attribute'') and by providing a parallel data structure for the additional information.
Additional surface-level methods (e.g. an architecture referring to a particular node also wants to have a convenient method to refer to it) can register their additional functions.

A workflow can combine defaults and particular consumers into one convenient surface.
Here, we show two simple examples of how such a convenience surface works, and for further details, we refer to the \cottax{} documentation.

In \cref{fig:interfaces_sellar}, the extended Sellar problem with an implicit discipline (of \cref{sec:architectures}) is shown in the \cottax{} interfaces in two different ways.
Functions can be decorated, giving just their output and reading the inputs from the signature, or can be converted to a \cottax{} function with possibly explicit paths into some kind of data structure.
The flat namespace is convenient for scripting and quick tests, while the explicit paths are useful to construct hierarchical model trees and data structures.
This can even be performed programmatically to arrive at a similar subsystem decomposition as performed in e.g. OpenMDAO.
Values in the graph declaration are converted to external inputs with an associated value (e.g. the bounds). The total declaration can be run in a simple manner, see \cref{fig:interfaces_architectures}.

\begin{figure*}[h]
    \input{figures/listing/sellar_three_listing.tex}
    \caption{Two different ways of specifying the problem: (a) in a flat namespace where every input and output just indexes into a Python dictionary and (b) indexing in an existing data structure using the builders (e.g. \texttt{V.design.y\_1}). }
    \label{fig:interfaces_sellar}
\end{figure*}

\begin{figure*}[h]
    \input{figures/listing/sellar_architectures.tex}
    \caption{Running the graph for two different architectures: default driver implementations are given, the starting values are given, the graph after the graph operations is created and it is run.
    Both architectures converge to the same result.}
    \label{fig:interfaces_architectures}
\end{figure*}

A similar interface is available for ODEs (using e.g. \texttt{flat.rate}); again, see the \cottax{} documentation for further details.
\subsection{Visualisation}\label{sec:visualisation}
All figures in this paper are drawn with the visualisation use-case in \cottax{}.
It can generate both the figures in \cref{sec:methods,sec:process} and interactive HTML versions that additionally step through a set of graph operations (see \cref{sec:process}).
\subsection{Architectures}\label{sec:architectures_cottax}
The architectures of \cref{sec:architectures} are implemented as a consumer (\cref{req:coordination}).
They are generic with respect to the conditional node and can be used both for algebraic relational and ODE solving and optimising (or even combining them).
MDF/IDF/SAND are directly implemented, including the possible rules for cutting (Gauss--Seidel with given order, Gauss--Seidel with minimal coupling, or Jacobi), and are used in the PROCESS results directly (\cref{sec:process}). 

They also do not require implementations for the functional and conditional nodes: the MDAO architecture can be defined with just the inputs and outputs of all functions and the conditions and unknowns of all conditional nodes defined.
\subsection{Execution}\label{sec:execution}
An execution layer is implemented using Python and can be executed eagerly for analysis workflows (using the Python interpreter as orchestration).

Most importantly, it can also be compiled and transformed directly using JAX \cite{jax2018github}.
During JAX tracing, the graph is traversed in Python and all array operations, functions, and the connections between them are recorded into a single program that JAX then compiles.
Tracing thus removes the graph-based core from the executed program entirely: the final program is equivalent to a manual ordering, satisfying \cref{req:overhead}.
See \cref{app:equal_jaxpr} for a worked example of this equivalence.
Trace time is very low (less than a second in \cref{sec:process}); compile time can be significant as it scales with the size of the program and has to be repeated for every change in the graph (but not for changing values; in \cref{sec:process} it is within 5 seconds in all cases).

The executable graph is then simply another JAX-compatible function and thus supports all JAX transformations. In particular, if all functions are in JAX, batching is trivial using the built-in \texttt{vmap} primitive: the entire executable graph can be executed in parallel. In \cref{sec:process}, thousands of design analyses and Jacobians are evaluated in parallel on the GPU (useful for e.g. the database generation or uncertainty quantification of \cref{req:overhead}).

Drivers can either be JAX-native drivers (e.g. Optimistix non-linear solvers \cite{optimistix2024} or Diffrax \cite{kidger2021on} for differential equations) or existing implementations in e.g. SciPy \cite{virtanen2020scipy} (which are handled through Python callbacks which are therefore not compiled in the same graph and do not necessarily support the JAX transformations).
Such drivers are handed their component subgraph, their required input data (e.g. initial guesses and tolerances), and the context (the boundary input variables of their component subgraph) --- they iterate (or integrate) the subgraph until their conditions are satisfied (nesting if required; gradient-based drivers may call the autodifferentiated program for a Jacobian). 

Drivers known ahead of time (i.e. specifically tailored to their associated function) can either be absorbed within that function or be assigned from the start.

Gradient-based optimisation or solving is supported through JAX's composable autodiff on the graph, satisfying \cref{req:gradients}.
Each functional node defining its vector-Jacobian product (VJP, for reverse mode) and Jacobian-vector product (JVP, for forward mode) is sufficient to yield an entirely differentiable graph.
This includes the drivers over the conditions: they can define a custom VJP/JVP using implicit differentiation over their solve using e.g. adjoint methods in Lineax \cite{lineax2023}, non-linear implicit differentiation in Optimistix, or implicit differentiation over optimisation problems \cite{blondel2022efficient}.
A VJP/JVP over the entire graph then automatically runs (nested) coupled adjoint or direct derivatives of the solves.

Non-native JAX functions can be wrapped using callbacks or e.g. in a more general manner \cite{hafner2025tesseract} --- this does introduce some overhead compared to pure-JAX programs. However, if the functions are sufficiently expensive (more than a few milliseconds), this overhead is negligible.
If derivatives are required, they should define a VJP or JVP, which is also possible using finite differences as fallbacks.
These finite differences can then be used as well for the residuals with an implicit VJP (i.e. using the adjoint method where the partial Jacobian is obtained from finite differences).
If models are wrapped in this way, JAX merely provides a chain rule to efficiently compose the VJP/JVPs for differentiability. 

Jacobian-colouring techniques are currently not implemented, but the dependency graph defines precisely the Jacobian sparsity pattern and could be exploited.
Alternatively, Jacobian-free Newton--Krylov solvers could be used, which only require the aforementioned JVP/VJP.

The current execution layer also does not exploit the model-level parallelism implicit in the graph (i.e. models not depending on each other's output that could be executed in parallel).
This does not map well to the JAX-parallelism model, but could later be implemented using e.g. callbacks or multi-device execution.

\clearpage

\section{PROCESS}\label{sec:process}
PROCESS \cite{kovari2014process,kovari2016process,lion2021general} is a fusion systems code with more than 140 interconnected models (differs per specific run) that are mostly scalar arithmetic.
This provides a compelling test case for the graph-based coordination and execution core: complex interdependencies with very fast single models.

This required PROCESS to be converted to a purely functional collection of models.
We converted, and in some cases split up, all models using LLM agents (Claude Sonnet 4.6 in most cases), testing each model against the existing PROCESS model extensively, including autodiff vs PROCESS's native finite differences. Some non-differentiable functions were smoothed slightly to obtain a differentiable version, taking care that this does not affect the result.

We then reproduced the input file optimisation runs to within reasonable tolerances ($\approx 10^{-3}$) using the \cottax{} JAX execution layer (and the MDAO architectures in \cref{sec:architectures}), which gives us confidence that it is correctly ported.

In this section, first, the collection of PROCESS models is sequenced for both a tokamak and a stellarator configuration, and the MDAO architectures are applied to this collection of models.
Then, some iteration, convergence, and performance characteristics are shown of the ported models including a comparison with the base PROCESS code for different MDAO architectures.
\subsection{Sequencing and MDAO architecture}
PROCESS contains models for both stellarators and tokamaks but also several device-agnostic models used for both.
We consider the \texttt{stellarator\_helias} (stellarator) and \texttt{low\_aspect\_ratio\_DEMO} (tokamak) example input files here --- these select a subset of the total models and pick which constraints to include and the design or iteration variables to vary.

\begin{figure}[bp]
    \centering
    \includegraphics{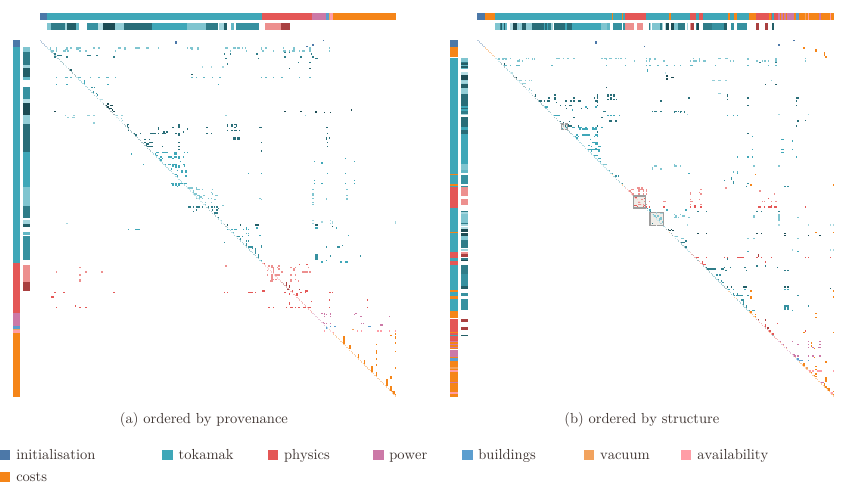}
    \caption{The tokamak (\texttt{low\_aspect\_ratio\_DEMO}) model collection (containing 242 models) ordered by provenance (a) and by a TOSCC (b), showing that the seemingly strongly cyclic nature of the model collection is an artefact of provenance-based ordering (PROCESS itself has a different ordering done by hand).
    The resulting cycles are small: see \cref{fig:process_blocks} for their precise definition.}
    \label{fig:process_1}
\end{figure}

In \cref{fig:process_1} the adjacency matrix of the tokamak is shown (similar to the figures in \cref{sec:methods} but without variable or node names in the graph and only one edge per coupling), coloured by the model provenance (e.g. physics, buildings, or coils) and ordered per provenance (without any graph operations to make it executable).
The non-sparsity below the diagonal indicates many coupling models in this particular run order.
However, after a TOSCC of this graph, shown in \cref{fig:process_1}(b), only very few cyclic components remain, and moreover, they are very small. 

\begin{figure}[tbp]
    \centering
    \includegraphics{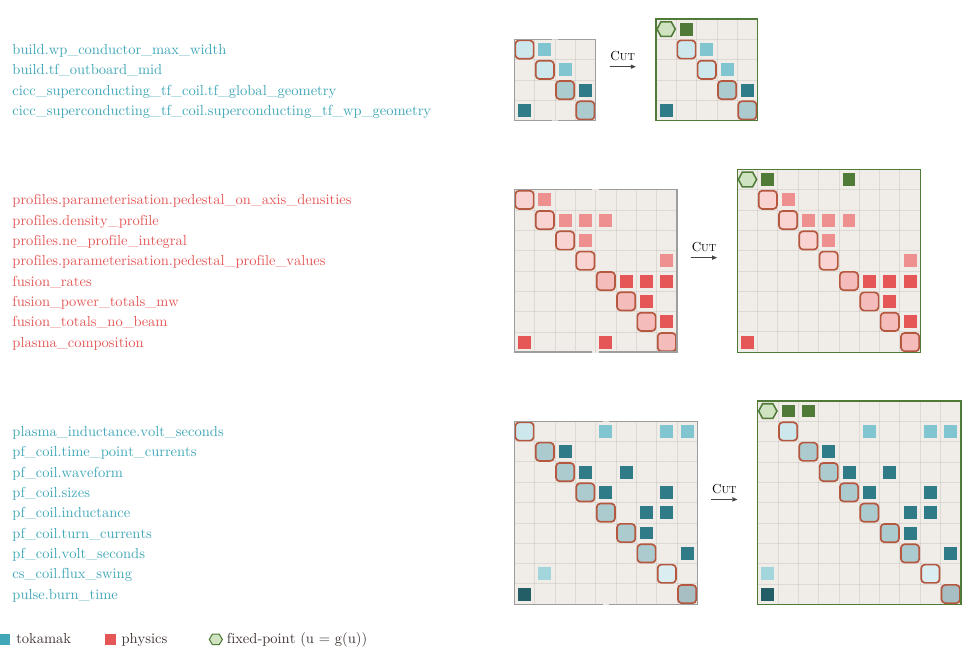}
    \caption{The strongly connected components (cycles) of \cref{fig:process_1} and their minimal Gauss--Seidel cut to turn them into functionally acyclic blocks (which makes the entire graph functionally acyclic).
    Provenance is indicated by colour.
    The top cycle is a winding-pack toroidal coil sizer, the middle a plasma composition solver, and the last a poloidal field coil sizer.}
    \label{fig:process_blocks}
\end{figure}
PROCESS itself runs a large fixed-point iteration over almost all its models --- this shows that with targeted ordering and solving, it can be reduced to only three small blocks of 4, 8, and 9 nodes (of a total of 242 here).
Moreover, they are sensible: a block over winding-pack sizing, a block over plasma composition, and a block over poloidal field coil sizing.

To recover PROCESS optimisation mode, an optimiser conditional node is added with the design variables as unknowns and the objective and constraints as conditions: in this case, the plant pulse burn time is maximised and nineteen design variables are used (see \cref{app:process_cases}).

\begin{figure}[tbp]
    \centering
    \includegraphics{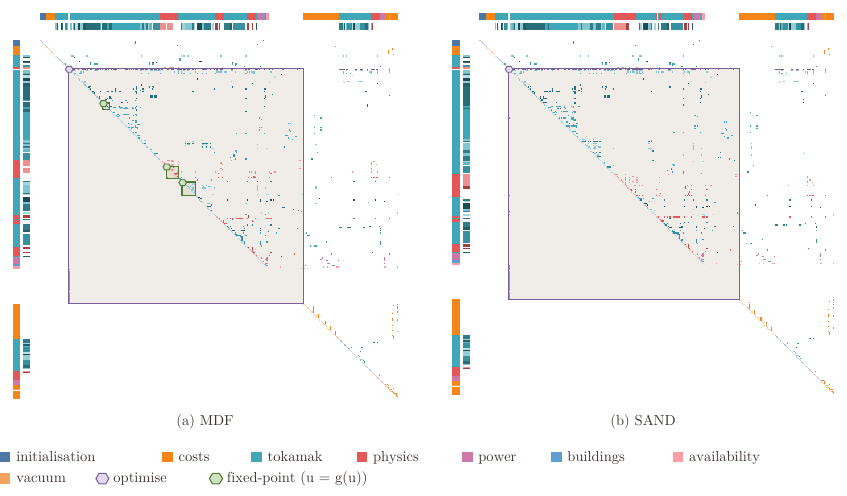}
    \caption{Optimisation architectures applied to the tokamak problem: an MDF architecture that resolves couplings internally (a) and a SAND architecture which resolves all couplings in the optimiser (b).
    The cycle induced by the optimiser does not include the entire graph: costing models are absent and are thus not iterated during the optimisation. }\label{fig:architectures_tok}
\end{figure}

Then, as discussed in \cref{sec:architectures}, an optimisation architecture can be defined in how the couplings in \cref{fig:process_blocks} are cut and then resolved.
First, everything is cut with minimal Gauss--Seidel (also in \cref{fig:process_blocks}).
Then, in \cref{fig:architectures_tok}, two choices are shown for combining and nesting: an MDF architecture and a SAND architecture (IDF has little meaning here because there are no implicit models and it would thus be the same as SAND). 

As the costing models do not feature in the objective nor in the constraints, they are absent from the optimiser cycle and are therefore not iterated during optimisation.
This was all fully automated: no hand-derived coupling or nesting was involved.

\begin{figure}[tbp]
    \centering
    \includegraphics{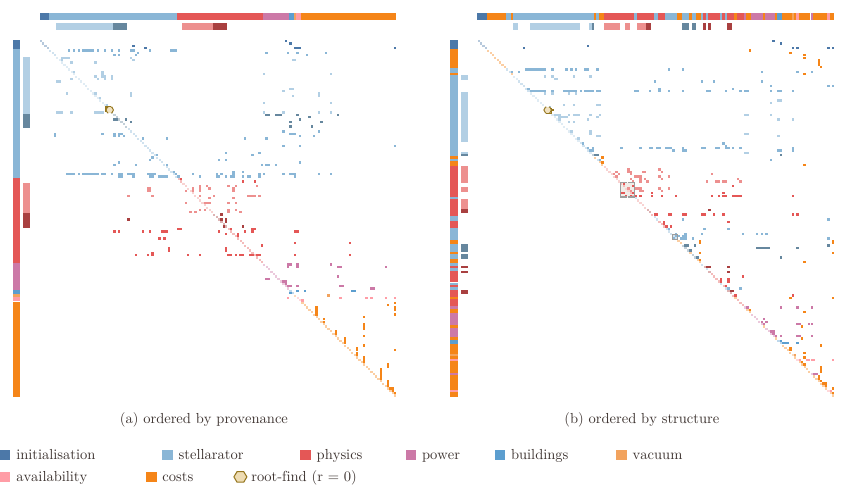}
    \caption{The stellarator model collection (149 models and one root-find conditional node) adjacency matrix shown ordered by provenance (a) and by TOSCC (b), after which it is almost entirely functionally acyclic.
    The remaining blocks are shown in \cref{fig:process_blocks_stellarator}. }
    \label{fig:process_stellarator}
\end{figure}

\begin{figure}[tbp]
    \centering
    \includegraphics{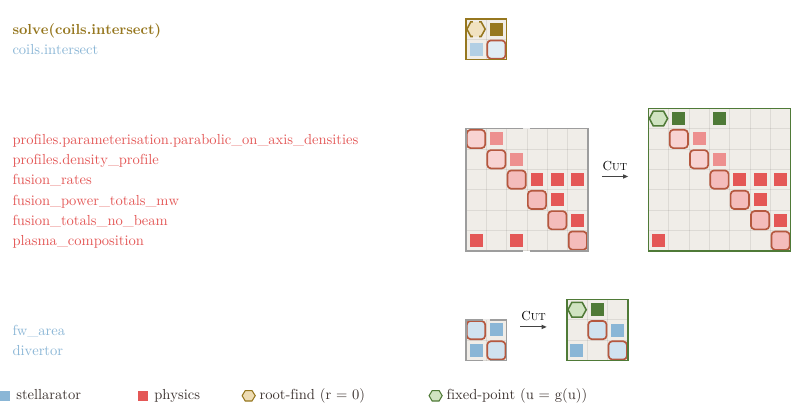}
    \caption{The strongly connected components of \cref{fig:process_stellarator} and their minimal Gauss--Seidel cut to turn them functionally acyclic.
    The top component is a coil intersect root find (an implicit model), the middle is the same plasma composition as for the tokamak in \cref{fig:process_blocks}, and the last a divertor-first-wall consistency cycle. }
    \label{fig:process_blocks_stellarator}
\end{figure}

\begin{figure}[tbp]
    \centering
    \includegraphics{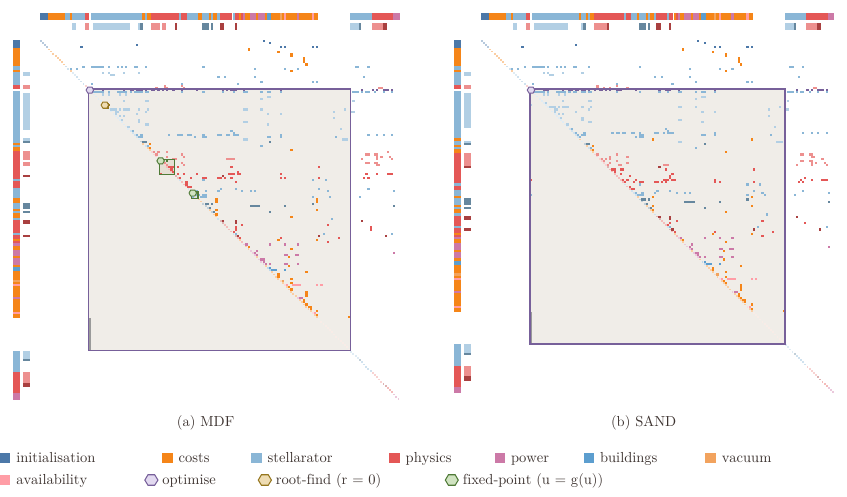}
    \caption{The stellarator model collection of \cref{fig:process_stellarator}, cut as indicated by \cref{fig:process_blocks_stellarator} with an MDF (a) and SAND (b) architecture. }
    \label{fig:architectures_stellarator}
\end{figure}

This exact same procedure is shown for the stellarator in \cref{fig:process_stellarator,fig:process_blocks_stellarator,fig:architectures_stellarator}, with the same conclusion: it is almost entirely acyclic and only some small blocks are actually coupled. 

These two configurations share only 70 models (out of 242 and 149 models) and have different objective functions and parametrisations (see \cref{app:process_cases}), showing that the coordination and automated MDAO architectures support an arbitrary model collection and parametrisation (\cref{req:selection,req:coordination}).

All diagrams are also available in HTML form for more detailed inspection at \url{https://timobogaarts.github.io/process-dsm}.

\clearpage
\subsection{Timing and results}\label{sec:process_timing}
Seven configurations (all shown explicitly in \cref{app:process_cases}) are then run and compared with the PROCESS results in terms of iterations and timing on the CPU (AMD EPYC 9654, Snellius Genoa) in \cref{tab:architectures}.

The Jacobian of the objective function is computed with automatic differentiation in all configurations; the nested solvers are differentiated using implicit differentiation provided by Optimistix \cite{optimistix2024} (\cref{req:gradients}).
The evaluation is one to two orders of magnitude faster and the \cottax{} Jacobian computation is two to three orders of magnitude faster than the finite differences used in base PROCESS. To make a fair comparison between base PROCESS and the \cottax{} port, the model columns in \cref{tab:architectures} show the time spent within the model during the optimisation --- excluding the overhead from the driver (VMCON or SLSQP) remaining in plain Python\footnote{To the best of our knowledge, there currently is no JAX-compatible constrained optimisation algorithm available that converges reliably with the problems shown here; it would be a useful addition to the ecosystem allowing JAX transformations on optimisers. The overhead of the \cottax{} wrapping is near-zero; the time difference is due to driver-internal time, and is significantly more in VMCON (Python-based) than in SLSQP (Fortran-based).}. The total model time is two to three orders of magnitude less in \cottax{}.

The ported functions converge to (within reasonable tolerance) the same optimum as PROCESS in all configurations.
Furthermore, the MDF (which is the most similar to PROCESS's own architecture) converges in the same or a smaller number of iterations (using VMCON). The table also motivates \cref{req:flexible}: no architecture outperforms every other in all cases. 

\begin{table*}[htbp]
    \centering\scriptsize
    \caption{Iterations and timings of \cottax{} under each MDAO architecture, beside PROCESS itself, for the seven regression-suite configurations of \cref{app:process_cases}: five optimisation configurations (top) and two evaluation configurations (bottom).
    IDF is listed only for the two stellarator configurations, the only ones with an implicit model.
    Each row gives the size of the outer problem (number of design variables $n$, equality constraints $m_\mathrm{eq}$ and inequality constraints $m_\mathrm{ineq}$), the number of graph nodes within the optimiser (excluding the nodes not in the optimiser cycle), and the number of iterations per driver (\checkmark: converged).
    The optimisation drivers are VMCON~\cite{crane1980vmcon} and SLSQP~\cite{kraft1988software}; for evaluation, the port uses a differentiable Newton method~\cite{nocedal_numerical_2006,optimistix2024} and PROCESS uses Powell's hybrid method~\cite{powell1970hybrid,virtanen2020scipy}.
    The eval and jac columns give the wall time of one evaluation and one Jacobian; each driver's time column gives the wall time of the whole solve, and the model column the part of it spent in the models (the difference is the driver-internal computation).
    $f^*$ is the objective at the optimum.}
    \label{tab:architectures}
    \setlength{\aboverulesep}{0pt}\setlength{\belowrulesep}{0pt}
\begin{tabular}{l@{\hspace{6pt}\vrule width 0.8pt\hspace{6pt}}rrrr@{\hspace{4pt}\vrule width 0.2pt\hspace{4pt}}rr@{\hspace{4pt}\vrule width 0.2pt\hspace{4pt}}rrrrrr@{\hspace{4pt}\vrule width 0.2pt\hspace{4pt}}r}
\toprule
\rule{0pt}{2.4ex} & & & & & \multicolumn{2}{c@{\hspace{4pt}\vrule width 0.2pt\hspace{4pt}}}{iterations} & \multicolumn{6}{c@{\hspace{4pt}\vrule width 0.2pt\hspace{4pt}}}{wall time (ms)} & \\
Optimisation configuration & $n$ & $m_\mathrm{eq}$ & $m_\mathrm{ineq}$ & nodes & VMCON & SLSQP & eval & jac & VMCON & model & SLSQP & model & $f^*$ \\[0.3ex]
\midrule
\rule{0pt}{2.4ex}\texttt{stellarator\_helias} & & & & & & & & & & & & & \\
cottax MDF & 8 & 2 & 12 & 122 & \checkmark\ 40 & \checkmark\ 26 & 0.24 & 0.51 & 460 & 100 & 79 & 55 & 1.218 \\
cottax IDF & 11 & 5 & 12 & 120 & \checkmark\ 47 & \checkmark\ 16 & 0.24 & 0.54 & 550 & 110 & 53 & 29 & 1.218 \\
cottax SAND & 12 & 6 & 12 & 119 & \checkmark\ 23 & \checkmark\ 16 & 0.26 & 0.56 & 290 & 56 & 52 & 29 & 1.218 \\
PROCESS & 8 & 2 & 12 & -- & \checkmark\ 82 & -- & 41 & 1100 & 180000 & 180000 & -- & -- & 1.215 \\[0.3ex]
\midrule
\rule{0pt}{2.4ex}\texttt{helias\_5b} & & & & & & & & & & & & & \\
cottax MDF & 3 & 2 & 2 & 73 & \checkmark\ 4 & \checkmark\ 4 & 0.23 & 0.30 & 54 & 6.9 & 23 & 6.1 & 0.7642 \\
cottax IDF & 6 & 5 & 2 & 71 & \checkmark\ 4 & \checkmark\ 4 & 0.23 & 0.39 & 56 & 6.7 & 24 & 6.4 & 0.7642 \\
cottax SAND & 6 & 5 & 2 & 71 & \checkmark\ 4 & \checkmark\ 4 & 0.23 & 0.41 & 56 & 7.5 & 24 & 6.5 & 0.7642 \\
PROCESS & 3 & 2 & 2 & -- & \checkmark\ 4 & -- & 44 & 440 & 3400 & 3100 & -- & -- & 0.7635 \\[0.3ex]
\midrule
\rule{0pt}{2.4ex}\texttt{large\_tokamak\_nof} & & & & & & & & & & & & & \\
cottax MDF & 20 & 3 & 23 & 177 & \checkmark\ 8 & \checkmark\ 7 & 0.38 & 2.2 & 160 & 51 & 59 & 30 & 1.600 \\
cottax SAND & 26 & 9 & 23 & 173 & \checkmark\ 9 & \checkmark\ 9 & 0.37 & 2.2 & 190 & 67 & 77 & 50 & 1.600 \\
PROCESS & 20 & 3 & 23 & -- & \checkmark\ 8 & -- & 21 & 1300 & 23000 & 22000 & -- & -- & 1.600 \\[0.3ex]
\midrule
\rule{0pt}{2.4ex}\texttt{low\_aspect\_ratio\_DEMO} & & & & & & & & & & & & & \\
cottax MDF & 19 & 4 & 21 & 178 & \checkmark\ 11 & \checkmark\ 11 & 1.1 & 4.8 & 270 & 130 & 120 & 83 & -0.4063 \\
cottax SAND & 24 & 9 & 21 & 175 & \checkmark\ 19 & \checkmark\ 21 & 1.1 & 4.7 & 470 & 240 & 210 & 180 & -0.4063 \\
PROCESS & 19 & 4 & 21 & -- & \checkmark\ 16 & -- & 27 & 1800 & 49000 & 48000 & -- & -- & -0.4063 \\[0.3ex]
\midrule
\rule{0pt}{2.4ex}\texttt{st\_regression} & & & & & & & & & & & & & \\
cottax MDF & 14 & 3 & 15 & 154 & \checkmark\ 10 & \checkmark\ 11 & 0.35 & 1.1 & 160 & 41 & 55 & 30 & -16.59 \\
cottax SAND & 17 & 6 & 15 & 152 & \checkmark\ 25 & \checkmark\ 81 & 0.33 & 1.2 & 400 & 130 & 380 & 340 & -16.59 \\
PROCESS & 14 & 3 & 15 & -- & \checkmark\ 10 & -- & 27 & 1200 & 25000 & 24000 & -- & -- & -16.59 \\[0.3ex]
\midrule\midrule
\rule{0pt}{2.4ex}Evaluation configuration & $n$ & $m_\mathrm{eq}$ & $m_\mathrm{ineq}$ & nodes & Newton & Powell & eval & jac & Newton & model & Powell & \multicolumn{1}{r@{\hspace{8.2pt}}}{model} & \\[0.3ex]
\midrule
\rule{0pt}{2.4ex}\texttt{large\_tokamak\_eval} & & & & & & & & & & & & \multicolumn{1}{r@{\hspace{8.2pt}}}{} & \\
cottax MDF & 2 & 2 & 0 & 34 & \checkmark\ 3 & -- & 0.17 & 0.23 & 11 & -- & -- & \multicolumn{1}{r@{\hspace{8.2pt}}}{--} & \\
PROCESS & 2 & 2 & 0 & -- & -- & \checkmark\ 10 & 21 & 150 & -- & -- & 740 & \multicolumn{1}{r@{\hspace{8.2pt}}}{280} & \\[0.3ex]
\midrule
\rule{0pt}{2.4ex}\texttt{spherical\_tokamak\_eval} & & & & & & & & & & & & \multicolumn{1}{r@{\hspace{8.2pt}}}{} & \\
cottax MDF & 3 & 3 & 0 & 38 & \checkmark\ 2 & -- & 0.17 & 0.27 & 9.7 & -- & -- & \multicolumn{1}{r@{\hspace{8.2pt}}}{--} & \\
PROCESS & 3 & 3 & 0 & -- & -- & \checkmark\ 10 & 24 & 250 & -- & -- & 820 & \multicolumn{1}{r@{\hspace{8.2pt}}}{300} & \\[0.3ex]
\bottomrule
\end{tabular}

\end{table*}

\begin{figure*}[htbp]
    \centering
    \includegraphics{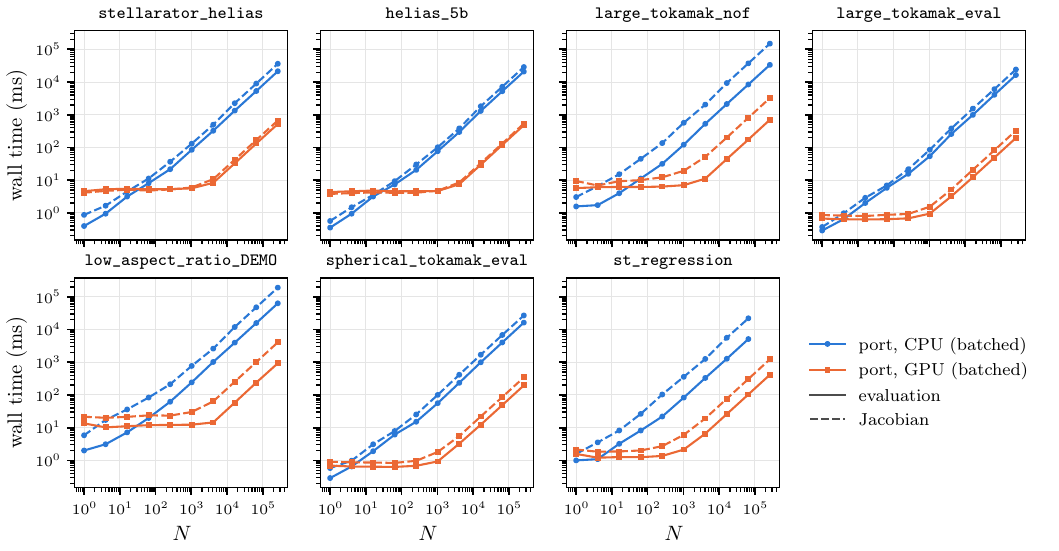}
    \caption{Wall times for $N$ evaluations of the MDA within the MDF structure (i.e. resolving the coupling but not optimising) for all seven configurations for the batched port (using \texttt{vmap}) using CPU (blue circle, AMD EPYC 9654) and GPU (orange square, NVIDIA H100).
    Solid lines are evaluations, dashed lines Jacobians.}
    \label{fig:batching}
\end{figure*}

The analysis can be batched arbitrarily on the GPU: although this is not necessarily an optimal workload for the GPU, the trivial parallelism can yield significant speedups for a large number of batched evaluations. 
This could be useful for uncertainty quantification or database generation, since each analysis is in principle a consistent analysis. 
In \cref{fig:batching} this batching is shown for an AMD EPYC 9654 CPU and an NVIDIA H100 GPU, showing that up to thousands of MDAs can be computed in the same time as a single evaluation on the GPU. 
Furthermore, for the stellarator, over two hundred thousand MDA Jacobians can be computed per second.

The computational speedup can in principle be obtained from a combination of four effects: compilation of all models, compilation of the graph, autodiff versus finite differences, and MDAO architecture changes. Finite differences have been tested using \cottax{} and are almost identical for execution speed (as these problems are close to square and with a relatively low number of design variables, forward mode was used which has similar computational costs; reverse mode is also supported) and convergence (within a few iterations of automatic differentiation).
To somewhat isolate the effect of the removal of the \cottax{} machinery from the evaluation during compilation, it is also run in interpreted mode for \texttt{stellarator\_helias}, compiling each model individually while not compiling \cottax{} execution itself (which thus runs in interpreted Python). 
This results in an evaluation of around 10 ms --- compilation of the graph orchestration (i.e. removing per-model JAX dispatch and \cottax{} interpreted Python graph traversal) provides a large speedup ($40 \times$) in this case (\cref{req:overhead}). The remaining 30 ms difference compared to PROCESS is a combination of architecture changes and model-level JIT. Other configurations have similar absolute timing differences of around 10 ms between interpreted and compiled modes.

\subsection{Summary}
The systems code PROCESS has been converted to a purely functional form that contains only the models (physics, engineering, and costing) themselves --- not the solvers or any other form of coordination.

These models have been wrapped in \cottax{} and sequenced, showing that only three small components ($\leq 9$ models) remain. 
For seven configurations (different models, parametrisations, and problem formulations, with less than half of the models shared between tokamaks and stellarators), the automated MDAO architecture definition runs and converges to the same values for two to three different MDAO architectures, demonstrating the required arbitrary design problems (\cref{req:selection}) and the automated coordination (\cref{req:coordination}).

No single architecture is in all cases the best-performing, motivating the MDAO architecture flexibility (\cref{req:flexible}).

Differentiation, required for the gradient-based optimisation of the drivers, is performed using automatic differentiation and implicit differentiation, demonstrating \cref{req:gradients} for PROCESS.

The model time (excluding driver-internal overhead and computations) of the \cottax{} port is two to three orders of magnitude faster for the optimisation configurations, and for \texttt{stellarator\_helias} a significant part of this difference is shown to be gained by removing per-model dispatch and graph-traversal overhead (\cref{req:overhead}).
This demonstrates graph-based flexibility and computational efficiency for a large number of low-fidelity models using \cottax{}.

\section{Discussion}\label{sec:discussion}
Fusion reactor design, and in particular stellarator design, requires integrated analysis and optimisation using increasingly high-fidelity models. To keep the future stellarator MDAO framework flexible and modular, we proposed a generic coordination and execution core which addresses the requirements given in \cref{sec:requirements} with a two-layered approach. 

The graph-based coordination layer aims to satisfy \cref{req:selection,req:coordination,req:flexible}, while the JAX-based execution layer addresses \cref{req:gradients,req:overhead}. We will discuss each requirement and how it is addressed here and provide limitations and future work. 

\subsection{Design problem agnostic}
\Cref{req:selection} is addressed by using functionally pure models in a generalised dependency graph. Parametrisations are also a node in the graph (e.g. a function that determines the analysis geometry as a function of major radius). A different problem formulation is an optimiser node with different unknowns; e.g. the optimiser determining a major radius or a magnetic field strength. The generality of this dependency graph ensures that no particular selection is enforced.

This does require a functionally pure set of models and parametrisations. For existing design frameworks this may require porting (as in PROCESS), but for future development, we would argue this is good practice in any case. By forcing explicit interfaces between models of subsystems it removes hidden dependencies that are also difficult to order, coordinate, and interpret manually.

\subsection{Automated coordination}
The standard MDAO architecture recipes of \cref{sec:architectures,sec:architectures_cottax} enable automated coordination for arbitrary design problems; for an acyclic set of models the topological ordering alone enables this. This has been shown concretely in the PROCESS results: for different model collections --- the tokamak and stellarator have vastly different model sets --- and various problem formulations (different design variables, objectives, and constraints) the MDAO architecture can be defined and run automatically.

The implemented monolithic MDAO architectures might not be the most efficient (see also \cref{req:flexible}), and additional architectures might be desirable as standard recipes. In particular, no distributed architectures \cite{tosserams_classification_2009} have been implemented. We propose they can however be obtained by sequences of the graph operations of \cref{sec:workflow} with additional functional nodes and conditional nodes. 

\subsection{Flexible MDAO architectures}
The graph operations of \cref{sec:workflow} can be used to define a highly customised MDAO architecture with arbitrarily deeply nested solves or optimisers. 

The additional solver flexibility enabled by the graph operations is only relevant for cases where the standard MDAO architectures do not yield sufficient performance.
This has not been shown in this paper yet: all applied MDAO architectures are one of the standard approaches. However, \cref{tab:architectures} does demonstrate that the difference in performance can be significant between architectures. 

Also, while the graph operations allow for customised MDAO architecture, there currently is no built-in check that guarantees the solved design optimisation problem is equal to the desired one. A possible future avenue to obtain such a check is to consider the fundamental problem graph \cite{van2019agile} and from there restrict the graph operations to a solution-set-preserving subset.

\subsection{Gradient computation}
JAX provides a wide variety of transformations on compatible Python functions, including gradients in forward or reverse mode. As the \cottax{} graph is such a function, it can be differentiated easily, including deeply nested solves using implicit differentiation techniques available in the JAX ecosystem. 

The efficiency of the algorithmic and implicit differentiation of a \cottax{} graph has to be shown for a large set of higher-fidelity models. PROCESS uses mostly scalar arithmetic and as such does not provide a sufficient test for this requirement. 
JAX is also already used for differentiable higher-fidelity simulations in fusion, e.g. DESC~\cite{dudt2020desc} and TORAX~\cite{torax2024arxiv}. Also, a to-be-published blanket optimisation with 200 design variables containing a higher-fidelity (over a billion degrees of freedom) differentiable neutronics model had no measurable difference in computational efficiency compared to our previous handwritten implementation (which is unsurprising given that the resulting compiled object does not contain \cottax{} machinery, see \cref{app:equal_jaxpr}).

Wrapping of non-JAX models (with possibly finite differences for gradients) has been tested (e.g. the PROCESS optimisers of \cref{sec:process_timing} are wrapped). In principle, with callbacks, this procedure would be similar to Python orchestration (e.g. OpenMDAO and GEMSEO), but no large-scale optimisations with opaque closed-source models have been attempted yet. In addition, using non-JAX differentiable components \cite{hafner2025tesseract} for opaque models that are not in JAX but can provide derivatives would be an interesting option: \cottax{} only requires explicit outputs and inputs and thus would natively support this.

Besides testing on a large set of higher-fidelity models and large-scale optimisations with opaque models, the development of a JAX-compatible constrained optimiser that can be implicitly differentiated and batched would be useful to reduce overhead and enable applications such as sensitivity analysis. Finally, inclusion of a MAUD-like \cite{Hwang2018a} derivative computation option could also be interesting.

\subsection{Low overhead}
The just-in-time compilation of the \cottax{} graph eliminates runtime overhead (\cref{sec:process_timing}) and enables batching of the same graph (\cref{fig:batching}).

Compilation overhead is relatively small in the PROCESS case (less than five seconds), but can be significant (tens of seconds) for very large and deeply nested graphs (thousands of nodes and more than four nesting layers). This can be worked around by e.g. exporting (precompiling) the function or using callbacks to particularly compile-intensive subprograms.

JAX does not natively support parallel execution of different models; callbacks could be used to parallelise this in the future, or a different execution layer could be used (see \cref{sec:further_outlook}).

Overhead reduction of a coordination and execution core is mostly relevant for cases with a relatively large number of models and relatively low computational time.
PROCESS is precisely this: it would be interesting to benchmark this overhead reduction in other applications.

\subsection{Further outlook}\label{sec:further_outlook}

As mentioned, the JAX execution layer has to be shown to be compatible with a larger set of higher-fidelity models. If it does not scale well in that regime, an alternative would be to develop a different execution layer while still using the \cottax{} coordination; the coordination does not depend on an implementation type (e.g. Julia functions or even repository locations are supported) and can even be performed without any implementation defined (\cref{sec:architectures_cottax}). Examples of possible execution layers are Julia (which could be an interesting option due to its relatively mature differentiable simulation ecosystem) or OpenMDAO. Exporting to e.g. CMDOWS \cite{van2018cmdows} could enable easier integration. 

The experimental \texttt{ode} conditional node type has to be tested more extensively and could be used for mixed algebraic / ODE usage for e.g. transport / integrated modelling, optimal control, or other time-dependent analyses.

A graph-based approach is useful beyond execution; it provides insight into model coupling and the required solves and can automatically define an MDAO architecture without even considering the function implementations. 
Insight into which models are coupled could be used to develop strongly coupled solves where needed using specialised software such as MOOSE~\cite{permann2020moose}. The reach of the graph (which variables are affected by which other) can also be used to determine valid design variables or unknowns for root finds. 
Finally, the \cottax{} graph could be integrated with model-based systems engineering, which also extensively uses graphs to similarly analyse design dependencies.

The stellarator MDAO framework requires a large amount of development still: models need to be developed for all relevant subsystems, a shared parametrised design ontology needs to be built (which these models have to be able to interface in) and these models and their coupled solves need to be verified and validated. 

Due to the arbitrary model collection support and flexible MDAO architecture definition, \cottax{} is not specific to stellarator MDAO or even fusion in general. 
The graph-based executability checks are general and could even be adapted to existing MDAO platforms such as OpenMDAO and GEMSEO.

\section{Conclusions}\label{sec:conclusions}
Motivated by the need for design optimisation including both high-fidelity models and many disciplines in nuclear fusion reactor design, we present the generic coordination and execution core \cottax{} that enables flexible (re)configuration and execution of multidisciplinary design analysis and optimisation (MDAO) architectures.

To achieve high flexibility, a graph-based definition of MDAO architectures is adopted.
In particular, we develop a method to control the solver structure on the level of individual iteration loops using a few elementary graph operations.
Standard monolithic MDAO architectures can be automatically assigned for arbitrary design problems. 
Furthermore, the general representation of both solvers and optimisers as iteration cycles with a condition to be satisfied allows future hybrid/custom approaches to also customise the MDAO architecture as desired, extending the flexibility of other graph-based approaches.

To obtain high computational efficiency for cases with higher-fidelity models and/or many low-fidelity models, a JAX execution layer is adopted.
JAX features just-in-time compilation, which eliminates overhead from the graph machinery and model dispatch which may bottleneck cases with a high number of low-fidelity models. 
The JAX execution layer allows the transformations of JAX --- differentiation and batching --- to be used directly on the \cottax{} graph itself.
This allows low-level parallelisation (using \texttt{vmap}) of low-fidelity models for database creation or uncertainty quantification applications and implicit differentiation over arbitrarily nested solves.
High-fidelity models can be directly included if they are JAX-native or otherwise wrapped.
We consider the JAX execution layer to be a suitable approach for a future stellarator MDAO framework, though other approaches (e.g. Julia) may be coupled to the graph coordination layer in future as well if needed.

The PROCESS case study illustrates the potential of the combined approach, as the JAX execution layer and MDAO architecture together reduce the model time during optimisation by two to three orders of magnitude.

Though full integration of a larger number of high-fidelity and/or black-box models remains to be demonstrated, the proposed coordination and execution core \cottax{} paves the way for design studies using selections of both high- and low-fidelity models which can easily and efficiently be combined to address a variety of design problems.

\cottax{} is designed to be general, modular, and extensible, and we hope that the ideas contained in it find broader use.

\begin{acknowledgments}
We would like to thank the PROCESS team at UKAEA for supporting this work. Their modernisation efforts have been critical in enabling the application of \cottax{} to PROCESS. We hope to contribute elements of this work to support the ongoing development of the PROCESS code. 
\end{acknowledgments}

\clearpage
\appendix
\section{Mathematical description}\label{app:maths}
The mathematical description of the checks in \cottax{} is given here; starting from the definition of a place, moving to the definition of nodes, the graph and its validity checks, a runnable partition, and finally the executable graph.

These mathematical constructs are repeated almost verbatim in the Python package, and furthermore allow for a comparison with the existing graph-based theory in \cite{pate2014graph,van2019agile,bruggeman2026dynamic}.

\subsection{Places}\label{app:places}
A place $p= (\rho_0, \rho_1, \dots) \in P$ is an ordered sequence of segments $\rho_j$ (e.g. indexes in a nested hierarchical structure).
The following operation is defined for $a,b \in P$ (both places):
\begin{alignat}{2}
  \textbf{Incomparable:}\quad a,b\in P: \ \ a \incomp b \defiff a \not\sqsubseteq b \text{ and } b \not\sqsubseteq a
\end{alignat}
Intuitively, comparable means that $a$ is either containing (a prefix of) $b$ in the hierarchical structure or vice versa (e.g. $c.d$ and $c$ are comparable): i.e. $a \sqsubseteq b $ iff $a$ is a prefix of $b$.
Two places being comparable is the generalisation of two places being equal in a flat namespace: two nodes with one owning $a.b$ and the other owning $a$ would equally be a conflict. 

\subsection{Nodes}
A node $n$ in the dependency graph is a set of inputs $I_n = \{p_j\}$ and outputs $O_n = \{p_k\}$, which are all simply a place:
\[ n = (I_n, O_n) \]

The following validity checks are enforced for every node $n \in N$:

\begin{checks}
    \item \label{cond:node_no_overlap} \textbf{No node owns something twice}:
    \[ o \incomp m \ \  \forall  \ o, m  \ \in O_n,\ o \neq m \]
    \item \label{cond:node_no_self_loop} \textbf{No node reads what it writes}:
    \[ o \incomp i \ \ \forall \ o \in O_n,\  i \in I_n\]
\end{checks}

\subsection{Graph}
A graph $G = (N)$ is constructed of valid nodes $N=\{n_i\}$ (it does not need to know about conditional or functional nodes to SCC) with edges $E$ directly implied by the nodes sharing inputs or outputs:
\begin{equation}
    E = \{ (n_i, n_j) \;:\; \exists\, o \in O_i,\ p \in I_j \ \text{ with }\ o = p \}.
\end{equation}
This has to satisfy the graph-validity checks:
\begin{checks}
    \item \label{cond:graph_one_producer} \textbf{One producer per place}: owned places are incomparable.
       \[ o \incomp m \ \ \forall\ n_i \neq n_j,\ o \in O_i,\ m \in O_j \]
    \item \label{cond:graph_exact_read} \textbf{Exact reads}: a read is exactly an owned place; no partial reads.
    \[ o \incomp p \ \text{ or }\ o = p \ \ \forall\ n_i, n_j,\ o \in O_i,\ p \in I_j \]
\end{checks}
The latter check is not entirely required: $a.b$ could be read if $a$ is owned, but for simplicity this is not implemented yet: simply pass the full $a$ instead (or make a projection-type node which extracts it).

Two nodes are $p \sim_G q$ iff they are reachable from each other through edges of the graph $G$ in both directions ($p \to q$ and $q \to p$ are paths in $G$, with path length $\geq 0$).
A \textit{strongly connected component} is then the equivalence class (grouping all nodes which are $\sim_G$):
\[[n]_G = \{ \, k \in N : k\sim_G n \, \}\]
In other words, $p \sim_G  q$ intuitively means that $p$ and $q$ are in the same strongly connected component of the graph $G$. 

By definition, the graph obtained by condensing the strongly connected components (taking each component as a single node) is acyclic (if there is a cycle between them, the nodes would be in the same component instead of two) and can be topologically sorted (TOSCC in the main text).

Acyclic means that each strongly connected component has size 1. 

\subsection{Node kinds}\label{sec:node_kinds}
Two types of nodes are defined:
\begin{alignat*}{2}
    \textbf{Functional node:}\quad & n = (I_n,O_n) && \quad n \in N_F\\ 
    \textbf{Conditional node:}\quad & n = (I_n,O_n, U_n, C_n) && \quad n \in N_C
\end{alignat*}
where $U_n=\{p_k\}$ are the unknowns and $C_n =\{p_l\}$ are the variables participating in a condition.
Thus, $N = N_F \sqcup N_C$ (a node is either conditional or functional).
These have the following validity checks:  

\begin{checks}
    \item \label{cond:func_has_outputs} \textbf{$\bm{N_F}$ has outputs}:
    \[ O_n \neq \emptyset \ \ \forall\ n \in N_F \]
    \item \label{cond:cond_unknowns_owned} \textbf{$\bm{N_C}$ outputs contain all unknowns}:
    \[ U_n \subseteq O_n \ \ \forall\ n \in N_C \]
    \item \label{cond:cond_conditions_read} \textbf{$\bm{N_C}$ inputs contain all conditions}:
    \[ C_n \subseteq I_n \ \ \forall\ n \in N_C \]
    \item \label{cond:cond_unknowns_empty} \textbf{$\bm{N_C}$ cannot be both without unknowns and without conditions}:
    \[ U_n \neq \emptyset \ \text{ or }\ C_n \neq \emptyset \ \ \forall\ n \in N_C \]
\end{checks}
A conditional node has additional inputs and outputs beyond its conditions and unknowns because a driver implementation might use those; the above checks are by construction true in the implementation.

\subsection{Nested graph}
A nested graph $G_N = (G, W)$ is an extension of a valid graph $G$ that adds explicit nesting edges $W = \{(p, q)\}$ with $p,q \in N$; $p$ is nested within $q$. 

Define the \textit{set of problems} of the set $N$ as $N_P = \{\, n \in N_C : U_n \neq \emptyset \,\}$.
Furthermore, the directed graph $(N_P, W)$ can be defined, as well as the set $A(q)\subseteq N_P$ all problems that a problem $q$ is transitively nested in (all problems with a path from $q$ in $(N_P, W)$, except itself).

The following checks are then required for a valid nesting:
\begin{checks}    
    \item \label{cond:nesting_bounds} \textbf{Nesting edges refer to problems}: bare conditions (without unknowns) cannot be nested.
    \[W \subseteq N_P \times N_P\]
    \item \label{cond:nesting_one_outer} \textbf{One outer}: a problem is nested in at most one outer problem.
    \[(n,m), (n, m') \in W \implies m = m'\]
    \item \label{cond:nesting_no_loops} \textbf{No nesting loops}: cannot nest problems within themselves.
    \[ (N_P, W) \text{ is a forest}\]
    \item \label{cond:nesting_one_cycle} \textbf{Nesting ends are on a cycle}: can only nest problems that are cyclic (using each other's outputs indirectly or directly).
    \[ p \sim_{G \setminus A(q)} q \ \ \forall \ (p,q)\in W\]
\end{checks}
The latter check is somewhat more complex: it means that in the subgraph obtained by removing all problems $q$ is nested in, the remaining inner ($p$) and outer problem ($q$) lie on the same component.

\subsection{Executable graph}
First, define the following: 
\begin{itemize}
    \item $S_p = [p]_{G \setminus A(p)} = \{\, k \in N \setminus A(p) : k \sim_{G \setminus A(p)} p \,\}$: the cycle of a problem $p$ with all its nesting ancestors removed (i.e. what it should iterate).
    \item $O(S) = \bigcup_{k\in S} O_k$ for a set of nodes $S$: all owned variables in that set $S$, and similarly for $I(S) = \bigcup_{k\in S} I_k$.
\end{itemize}

The driver contract is that it iterates $U_p \to C_p$ with the additional inputs and outputs used only at the start and at the end.
In other words:
\begin{itemize}
    \item \textbf{Read} $I_p \setminus C_p$: driver input data such as tolerances or initial guesses.
    \item \textbf{Iterate} $U_p \to C_p$: until conditions are met.
    \item \textbf{Output} $O_p \setminus U_p$: driver output data such as iteration history.
\end{itemize}
Additional data (e.g. inverse Jacobian approximations) is driver-internal and not in the graph. 

A nested graph is executable if additionally the following checks hold:

\begin{checks}
    \item \label{cond:no_bare_conditions} \textbf{No bare conditions}: all have unknowns.
     \[N_P = N_C\]
    \item \label{cond:func_acyclic} \textbf{No functional nodes have feedback}: only conditional nodes are allowed to have feedback.
     \[ G\setminus N_P \text{ is acyclic }\]
    \item \label{cond:conditions_on_cycle} \textbf{Conditions determined on the cycle}: a problem iterates something and every condition it holds is produced on that cycle.
    \[|S_p| > 1 \ \text{ and } \ C_p \subseteq O(S_p) \ \ \ \forall p \in N_P \]
    \item \label{cond:siblings_cannot_be_cyclical} \textbf{One problem per level}: nodes cannot be cyclic if they answer on the same nesting level (i.e. have the same upwards nesting edge, such that $A(q) = A(q')$).
    \[q \not\sim_{G \setminus A(q)} q' \quad \forall q \neq q'\in N_P \text{ with } \ A(q) = A(q')\]
    \item \label{cond:driver_data_not_on_cycle} \textbf{Driver data not produced or read on cycle}: the inputs of a driver (e.g. tolerance) cannot be produced on its cycle and its outputs (e.g. iteration history) cannot be read on its cycle.
    The driver iterates $U_p \to C_p$ only.
    \[(I_p \setminus C_p) \cap  O(S_p) = \emptyset  \ \ \ \forall p \in N_P \]
    \[(O_p \setminus U_p) \cap  I(S_p) = \emptyset  \ \ \ \forall p \in N_P \]
    
\end{checks}
Furthermore, for the execution layer of \cottax{}, the driver should be assigned and compatible with the problem; these are not graph checks but rather implementation.

In \cref{tab:check_mapping} the correspondence between the checks given in \cref{sec:methods} and the checks given in the appendix is given. The checks that have no version in the main text of \cref{sec:methods} are the definitions of the node kinds and the check that driver data are not produced or read on the cycle; these are assumed in the main text to be valid in all cases.

\begin{table}[htbp]
    \centering\small
    \caption{Correspondence between the checks of \cref{sec:methods} and the checks of this appendix.}\label{tab:check_mapping}
    \begin{tabular}{@{}llp{0.42\linewidth}@{}}
        \toprule
        \multicolumn{2}{@{}l}{Main text} & Appendix \\
        \midrule
        \ref{cond:one_producer}       & One producer per variable  & \ref{cond:node_no_overlap}, \ref{cond:graph_one_producer} \\
        \ref{cond:no_self_read}       & No self-reads              & \ref{cond:node_no_self_loop} \\
        \ref{cond:no_nesting_loops}   & No nesting loops           & \ref{cond:nesting_no_loops} \\
        \midrule
        \ref{cond:condition_unknowns} & Conditions determined      & \ref{cond:no_bare_conditions}, \ref{cond:conditions_on_cycle} \\
        \ref{cond:func_acyclic_2}     & Functionally acyclic       & \ref{cond:func_acyclic} \\
        \ref{cond:driver_assigned}    & Drivers assigned           & --- (implementation, not a graph check) \\
        \ref{cond:valid_nesting}      & Valid nesting              & \ref{cond:nesting_bounds}, \ref{cond:nesting_one_outer}, \ref{cond:nesting_one_cycle}, \ref{cond:siblings_cannot_be_cyclical} \\
        \midrule
        \multicolumn{2}{@{}l}{--- (appendix only)} & \ref{cond:graph_exact_read}, \ref{cond:func_has_outputs}, \ref{cond:cond_unknowns_owned}, \ref{cond:cond_conditions_read}, \ref{cond:cond_unknowns_empty}, \ref{cond:driver_data_not_on_cycle} \\
        \bottomrule
    \end{tabular}
\end{table}

\subsection{Comparison with existing graph-based framework}\label{app:kadmos}
The graph-based formulation of MDAO problems in KADMOS \cite{van2019agile} similarly has defined several checks that have to be passed in order to define an MDAO problem. 

First, it has a somewhat broader and different scope: it also concerns the earlier stages before a particular selection of models is chosen, has special roles for the outer problem formulation (with specifically named objective, constraints, and design variables in the graph checks), and is in general tailored for MDAO problem specification rather than simply being executable (as this work is).
This work tries to formalise from as little as possible how a dependency graph with conditional nodes looks and when it is valid, and argues for MDAO architectures to be built on top of that, rather than starting from the MDAO formulation itself.

Pate et al.~\cite{pate2014graph} and KADMOS \cite{van2019agile} both use bipartite graphs instead: this maps one-to-one to the definition in \cref{app:maths} however.
The main difference is that the conditional node extension is not explicit in KADMOS and therefore many checks are not expressible in KADMOS or vice versa.

Process graphs as in \cite{van2019agile} are not considered in our formulation: a TOSCC after having checked executability explicitly defines a valid process graph and thus there is nothing to check.
Furthermore, since we do not construct a repository connectivity graph (the collection of all possible models, which can have overlapping outputs), checks that reference it are not relevant here. 

\begin{table*}[htbp]
    \centering\scriptsize
    \caption{Checks from earlier graph-based MDAO formulations, grouped by graph type, beside the corresponding checks of this work.}\label{tab:kadmos_comparison}
    \setlength{\aboverulesep}{0pt}\setlength{\belowrulesep}{0pt}\renewcommand{\arraystretch}{1.15}
    \begin{tabular}{@{}l p{0.41\linewidth}@{\hskip10pt\vrule width 0.4pt\relax\hskip10pt}p{0.43\linewidth}@{}}
        \toprule
        & Pate et al.\ / KADMOS \cite{van2019agile} & This work \\
        \midrule
        \multicolumn{3}{@{}l}{\textit{Data graph} (base class of RCG, FPG, MDG)} \\
        1--2 & nodes are functions or variables; edges connect a variable and a function
            & --- (by construction: nodes are always functions or conditions: edges implied) \\
        \midrule
        \multicolumn{3}{@{}l}{\textit{Repository connectivity graph (RCG)}} \\
        --- & (no checks beyond base data graph) & ---\\
        \midrule
        \multicolumn{3}{@{}l}{\textit{Fundamental problem graph (FPG)}} \\
        1--2 & functions and variables are in the RCG, or instances/merges of RCG nodes
            & --- (no RCG) \\
        3 & no \emph{collisions} ($\delta^-(v)\le 1$); no \emph{circular} variables ($\mathrm{cir}(v)=0$); no isolated variables
            & \cref{cond:node_no_overlap,cond:graph_one_producer}; \cref{cond:node_no_self_loop}; (isolated variables not representable)\\
        4 & functions have outputs ($\delta^+>0$)
            & \cref{cond:func_has_outputs} \\
        5 & functions have a problem role
            & --- (follows from TOSCC) \\
        6 & design variables are inputs ($\delta^-=0$)
            & --- (no special role for optimiser)\\
        7 & quantities of interest have one producer ($\delta^-=1$)
            & \cref{cond:graph_one_producer} but we require for every node in the graph \\
        8 & objective and constraints are outputs ($\delta^+=0$)
            & --- only that they are read by a conditional node: \cref{cond:cond_conditions_read} \\
        9 & every node lies on a path to a QOI, objective or constraint
            & --- (trimming not relevant for executability) \\
        \midrule
        \multicolumn{3}{@{}l}{\textit{MDAO data graph (MDG)}} \\
        1 & contains all FPG nodes
            & --- \\
        2 & a single coordinator
            & --- (top level is $G\setminus N_P$, ordered by TOSCC, multiple are fine) \\
        3 & every node lies on a cycle through the coordinator
            & --- (no single top-level)\\
        4--6 & architecture roles of functions (optimiser, converger, \dots) and variables (initial guess, copy, final value)
            & --- (architecture is not enforced) \\        
        \midrule
        \multicolumn{3}{@{}l}{\textit{MDAO process graph (MPG)}} \\
        1 & nodes are the MDG functions
            & --- \\
        2--4 & starts and ends at the coordinator; every node on a cycle through it
            & --- (order follows from TOSCC and nesting) \\
        5 & a cycle with continuously increasing step numbers
            & --- (idem) \\
        \midrule
        \multicolumn{3}{@{}l}{\textit{Dynamic workflows} \cite{bruggeman2026dynamic}} \\
        & a design variable is defined at one workflow level only
            & \cref{cond:graph_one_producer} (unknowns owned once, across nesting) \\
        & a subworkflow is a function node holding its own graph
            & similar to \cref{cond:nesting_one_outer,cond:nesting_no_loops} \\
        & subworkflow inputs (incl.\ initial guesses) supplied by, outputs returned to, the parent
            & \cref{cond:driver_data_not_on_cycle} for initial guesses\\
        & switches activate one of several mutually exclusive branches
            & --- (graph fixed during execution) \\
        \midrule
        \multicolumn{3}{@{}l}{\textit{Not checked in KADMOS} (checks of this work)} \\
        & --- (coupling allowed in the FPG; resolved by a converger with copy variables in the MDG algorithm)
            & \cref{cond:func_acyclic} no functional feedback \\
        \midrule
        \multicolumn{3}{@{}l}{\textit{Not expressible in KADMOS} (checks of this work)} \\
        & --- (variable places are not hierarchical)
            & \cref{cond:graph_exact_read} exact reads: a read is exactly an owned place \\[0.5ex]
        & --- (no conditional nodes: drivers are function nodes with an architecture role, inserted by the architecture algorithms)
            & \cref{cond:cond_unknowns_owned} conditional nodes own their unknowns\newline
              \cref{cond:cond_unknowns_empty} conditional nodes have unknowns or conditions\newline
              \cref{cond:no_bare_conditions} every condition has unknowns\newline
              \cref{cond:conditions_on_cycle} conditions determined on the cycle \\[0.5ex]
        & --- (no nesting relation: loop nesting is implicit in process step numbers)
            & \cref{cond:nesting_bounds} only problems can be nested\newline
              \cref{cond:nesting_one_cycle} nested problems share a cycle with their outer problem\newline
              \cref{cond:siblings_cannot_be_cyclical} one problem per nesting level \\
        \bottomrule
    \end{tabular}
\end{table*}

\section{Equality of programs: manual and \cottax{}}\label{app:equal_jaxpr}
In this appendix, the same model is constructed using imperative programming and \cottax{}.
These will yield the exact same \texttt{jaxpr} (intermediate representation of JAX), which is then just-in-time compiled.
The same \texttt{jaxpr} yields the same computational graph.  

In general this is only identical up to the level at which the topological sort is unambiguous: parallel-executable nodes may be reordered arbitrarily and as such in general not be equal to a manually ordered structure --- this example is unambiguous by construction.
\begin{figure*}[h]
    \input{figures/listing/jaxpr_acyclic.tex}
    \caption{The same acyclic model ordered by hand (left) and declared in \cottax{} and ordered by the graph (right).
    Both trace to the identical \texttt{jaxpr}, printed at the bottom.}
    \label{fig:jaxpr_acyclic}
\end{figure*}

Driven blocks are somewhat more complicated: the driven subgraph is evaluated once more after converging to ensure consistency of the data structure with the final unknowns (the unknowns can be updated after the last value iteration --- values in the data structure are one step stale).
This ensures derivatives are computed accurately in all cases. 

In some cases, this may be less efficient than a hand-written solver loop, which, given the internals of the solver and details of the problem, could eliminate some of this recomputation in favour of reusing final iterate values.
Currently, in \cottax{}, for simplicity, such elimination is not performed: the cost is low compared to the iterative cost (one value iteration extra compared to tens to hundreds of iterations possibly including Jacobians) and evaluates values and derivatives correctly whatever the problem or driver\footnote{This does not necessarily include driver outputs: iteration history is inherently dependent on the iterative path, while the conditions / unknowns are not.
Such a history cannot be differentiated through the implicit function theorem; only by unrolling all iterations.
Since this is extremely memory-intensive in reverse mode, this will be set to zero in all cases. }

\section{Definition of PROCESS configurations}\label{app:process_cases}
All PROCESS configurations considered here are in the regression suite of the PROCESS GitHub.
The tables below give explicitly the iteration variables, objective (if present), and equality and inequality constraints. 
In the root-find (evaluation) configurations, only the equality constraints are solved; the listed inequalities are evaluated at the solution but not imposed.
In \cref{tab:problem-switches} the particular model switches are also shown for all seven configurations.

\begin{table*}[htbp]
\centering\scriptsize
\caption{Symbols used in the PROCESS configurations}  %
\label{tab:problem-symbols}
\setlength{\aboverulesep}{0pt}\setlength{\belowrulesep}{0pt}
\begin{tabular}{@{}p{\dimexpr 0.15\textwidth-7.550pt\relax}@{\hspace{10pt}}p{\dimexpr 0.35\textwidth-7.550pt\relax}@{\hspace{5pt}\vrule width 0.2pt\relax\hspace{5pt}}p{\dimexpr 0.15\textwidth-7.550pt\relax}@{\hspace{10pt}}p{\dimexpr 0.35\textwidth-7.550pt\relax}@{}}
\toprule
\rule{0pt}{2.6ex}\rule[-1.2ex]{0pt}{0pt}\raggedright symbol & \raggedright meaning & \raggedright symbol & \raggedright meaning \tabularnewline
\midrule
\rule{0pt}{2.6ex}\raggedright $B_\mathrm{t}(R_0)$ & \raggedright toroidal field on axis & \raggedright $P_\mathrm{fus}, P_\mathrm{net}$ & \raggedright fusion power, net electric power \tabularnewline
\raggedright $R_0$ & \raggedright plasma major radius & \raggedright $P_\mathrm{sep}, P_\mathrm{LH}$ & \raggedright power across the separatrix, L--H threshold power \tabularnewline
\raggedright $A$ & \raggedright aspect ratio & \raggedright $P_\mathrm{aux}$ & \raggedright injected auxiliary power \tabularnewline
\raggedright $\langle T_\mathrm{e}\rangle$ & \raggedright volume-averaged electron temperature & \raggedright $f_\mathrm{rad}$ & \raggedright radiated power fraction \tabularnewline
\raggedright $\langle n_\mathrm{e}\rangle$ & \raggedright volume-averaged electron density & \raggedright $q_\mathrm{n}$ & \raggedright neutron wall load \tabularnewline
\raggedright $n_\mathrm{e0}, n_\mathrm{e,ped}$ & \raggedright electron density on axis, at the pedestal & \raggedright $q_\mathrm{div}, q_\mathrm{FW,rad}$ & \raggedright divertor heat load, first-wall radiation load \tabularnewline
\raggedright $n_\mathrm{lim}$ & \raggedright density limit & \raggedright $B_\mathrm{max}$ & \raggedright peak field on the coil \tabularnewline
\raggedright $\langle n_{\alpha,\mathrm{th}}\rangle$ & \raggedright thermal alpha density & \raggedright $j_\mathrm{CS}, j_\mathrm{TF,WP}$ & \raggedright CS, TF winding pack current density \tabularnewline
\raggedright $n_\mathrm{Xe}$ & \raggedright xenon impurity density & \raggedright $j_\mathrm{crit}, j_\mathrm{quench}$ & \raggedright critical current density, quench (hot-spot) limit \tabularnewline
\raggedright $f_\mathrm{GW,ped}, f_\mathrm{GW,sep}$ & \raggedright Greenwald fraction of the pedestal, separatrix density & \raggedright EOF, BOP & \raggedright end of flat-top, beginning of pulse \tabularnewline
\raggedright $\langle\beta\rangle$ & \raggedright volume-averaged total beta & \raggedright $\sigma_\mathrm{TF,case}$, $\sigma_\mathrm{TF,conduit}$ & \raggedright TF coil case, conduit stress \tabularnewline
\raggedright $H$ & \raggedright confinement enhancement factor (hfact; $f_\mathrm{ren}$ in \cite{lion2021general}) & \raggedright $V_\mathrm{TF}, \tau_\mathrm{Q}$ & \raggedright TF quench dump voltage, discharge time \tabularnewline
\raggedright $q_{95}$ & \raggedright safety factor at the 95\% flux surface & \raggedright $\Delta T_\mathrm{TF}, \Delta T_\mathrm{CS}$ & \raggedright superconductor temperature margin \tabularnewline
\raggedright $f_\mathrm{NI}$ & \raggedright non-inductive fraction of the plasma current & \raggedright $I_\mathrm{turn}$ & \raggedright TF current per turn \tabularnewline
\raggedright $I_\mathrm{p}, I_\mathrm{rod}$ & \raggedright plasma current, centre-post current & \raggedright $f_\mathrm{Cu}$ & \raggedright copper fraction of the TF cable \tabularnewline
\raggedright $\tau_\alpha, \tau_\mathrm{E}$ & \raggedright alpha particle, energy confinement time & \raggedright $N_\mathrm{cycles,CS}$ & \raggedright achievable CS stress cycles \tabularnewline
\raggedright $Q$ & \raggedright fusion gain & \raggedright $d_\mathrm{min}, w_\mathrm{coil}$ & \raggedright minimal distance between coils, coil toroidal width \tabularnewline
\raggedright $t_\mathrm{burn}$ & \raggedright burn time (pulse length) & \raggedright $d_\mathrm{pc}, \textstyle\sum d$ & \raggedright plasma--coil distance, required radial build \tabularnewline
\rule[-1.2ex]{0pt}{0pt}\raggedright CoE & \raggedright cost of electricity & \raggedright $\sigma_\mathrm{VV,quench}$, $\sigma_\mathrm{CS,Tresca}$ & \raggedright vacuum vessel stress during a TF quench, CS Tresca stress \tabularnewline
\bottomrule
\end{tabular}
\end{table*}

\begin{table*}[htbp]
\centering\scriptsize
\caption{\texttt{stellarator\_helias}}  %
\label{tab:problem-stellarator-helias}
\setlength{\aboverulesep}{0pt}\setlength{\belowrulesep}{0pt}
\begin{tabular}{@{}p{\dimexpr 0.37\textwidth-7.650pt\relax}@{\hspace{5pt}\vrule width 0.2pt\relax\hspace{5pt}}p{\dimexpr 0.15\textwidth-7.650pt\relax}@{\hspace{5pt}\vrule width 0.2pt\relax\hspace{5pt}}p{\dimexpr 0.19\textwidth-7.650pt\relax}@{\hspace{5pt}\vrule width 0.2pt\relax\hspace{5pt}}p{\dimexpr 0.29\textwidth-7.650pt\relax}@{}}
\toprule
\rule{0pt}{2.6ex}\rule[-1.2ex]{0pt}{0pt}\raggedright design variables (\texttt{ixc}) & \raggedright objective & \raggedright equality constraints (\texttt{icc}) & \raggedright inequality constraints (\texttt{icc}) \tabularnewline
\midrule
\begin{tabular}[t]{@{}r@{\hspace{0.6em}}p{\dimexpr 0.37\textwidth-7.650pt\relax-2.6em}@{}}
2 & \raggedright \rule{0pt}{2.6ex}$B_\mathrm{t}(R_0)$ $\in$ [\num{4}, \num{10}] T \tabularnewline
3 & \raggedright $R_0$ $\in$ [\num{10}, \num{30}] m \tabularnewline
4 & \raggedright $\langle T_\mathrm{e}\rangle$ $\in$ [\num{3}, \num{15}] keV \tabularnewline
6 & \raggedright $\langle n_\mathrm{e}\rangle$ $\in$ [\num{3e+19}, \num{3e+20}] m$^{-3}$ \tabularnewline
10 & \raggedright $H$ $\in$ [\num{0.1}, \num{1.2}] \tabularnewline
56 & \raggedright $\tau_\mathrm{Q}$ $\in$ [\num{1}, \num{50}] s \tabularnewline
59 & \raggedright $f_\mathrm{Cu}$ $\in$ [\num{0.3}, \num{0.9}] \tabularnewline
109 & \raggedright $\langle n_{\alpha,\mathrm{th}}\rangle/\langle n_\mathrm{e}\rangle$ $\in$ [\num{0.0001}, \num{0.4}]\rule[-1.2ex]{0pt}{0pt} \tabularnewline
\end{tabular} &
\begin{tabular}[t]{@{}r@{\hspace{0.6em}}p{\dimexpr 0.15\textwidth-7.650pt\relax-2.6em}@{}}
6 & \raggedright \rule{0pt}{2.6ex}min CoE\rule[-1.2ex]{0pt}{0pt} \tabularnewline
\end{tabular} &
\begin{tabular}[t]{@{}r@{\hspace{0.6em}}p{\dimexpr 0.19\textwidth-7.650pt\relax-2.6em}@{}}
2 & \raggedright \rule{0pt}{2.6ex}Power balance \tabularnewline
16 & \raggedright $P_\mathrm{net} \ge\mathrm{min}$\rule[-1.2ex]{0pt}{0pt} \tabularnewline
\end{tabular} &
\begin{tabular}[t]{@{}r@{\hspace{0.6em}}p{\dimexpr 0.29\textwidth-7.650pt\relax-2.6em}@{}}
8 & \raggedright \rule{0pt}{2.6ex}$q_\mathrm{n} \le\mathrm{max}$ \tabularnewline
17 & \raggedright $f_\mathrm{rad} \le\mathrm{max}$ \tabularnewline
18 & \raggedright $q_\mathrm{div} \le\mathrm{max}$ \tabularnewline
24 & \raggedright $\langle\beta\rangle \le\mathrm{max}$ \tabularnewline
32 & \raggedright $\sigma_\mathrm{TF,conduit} \le\mathrm{max}$ \tabularnewline
34 & \raggedright $V_\mathrm{TF} \le\mathrm{max}$ \tabularnewline
35 & \raggedright $j_\mathrm{TF,WP} \le j_\mathrm{quench}$ \tabularnewline
62 & \raggedright $\tau_\alpha/\tau_\mathrm{E} \ge\mathrm{min}$ \tabularnewline
65 & \raggedright $\sigma_\mathrm{VV,quench} \le\mathrm{max}$ \tabularnewline
67 & \raggedright $q_\mathrm{FW,rad} \le\mathrm{max}$ \tabularnewline
82 & \raggedright $d_\mathrm{min} \ge w_\mathrm{coil}$ \tabularnewline
83 & \raggedright $d_\mathrm{pc} \ge \textstyle\sum d$\rule[-1.2ex]{0pt}{0pt} \tabularnewline
\end{tabular} \tabularnewline
\bottomrule
\end{tabular}
\end{table*}

\begin{table*}[htbp]
\centering\scriptsize
\caption{\texttt{helias\_5b}}  %
\label{tab:problem-helias-5b}
\setlength{\aboverulesep}{0pt}\setlength{\belowrulesep}{0pt}
\begin{tabular}{@{}p{\dimexpr 0.37\textwidth-7.650pt\relax}@{\hspace{5pt}\vrule width 0.2pt\relax\hspace{5pt}}p{\dimexpr 0.15\textwidth-7.650pt\relax}@{\hspace{5pt}\vrule width 0.2pt\relax\hspace{5pt}}p{\dimexpr 0.19\textwidth-7.650pt\relax}@{\hspace{5pt}\vrule width 0.2pt\relax\hspace{5pt}}p{\dimexpr 0.29\textwidth-7.650pt\relax}@{}}
\toprule
\rule{0pt}{2.6ex}\rule[-1.2ex]{0pt}{0pt}\raggedright design variables (\texttt{ixc}) & \raggedright objective & \raggedright equality constraints (\texttt{icc}) & \raggedright inequality constraints (\texttt{icc}) \tabularnewline
\midrule
\begin{tabular}[t]{@{}r@{\hspace{0.6em}}p{\dimexpr 0.37\textwidth-7.650pt\relax-2.6em}@{}}
4 & \raggedright \rule{0pt}{2.6ex}$\langle T_\mathrm{e}\rangle$ $\in$ [\num{4}, \num{25}] keV \tabularnewline
6 & \raggedright $\langle n_\mathrm{e}\rangle$ $\in$ [\num{3.005e+19}, \num{5.005e+20}] m$^{-3}$ \tabularnewline
10 & \raggedright $H$ $\in$ [\num{0.1}, \num{2}]\rule[-1.2ex]{0pt}{0pt} \tabularnewline
\end{tabular} &
\begin{tabular}[t]{@{}r@{\hspace{0.6em}}p{\dimexpr 0.15\textwidth-7.650pt\relax-2.6em}@{}}
7 & \raggedright \rule{0pt}{2.6ex}min capital cost\rule[-1.2ex]{0pt}{0pt} \tabularnewline
\end{tabular} &
\begin{tabular}[t]{@{}r@{\hspace{0.6em}}p{\dimexpr 0.19\textwidth-7.650pt\relax-2.6em}@{}}
2 & \raggedright \rule{0pt}{2.6ex}Power balance \tabularnewline
16 & \raggedright $P_\mathrm{net} \ge\mathrm{min}$\rule[-1.2ex]{0pt}{0pt} \tabularnewline
\end{tabular} &
\begin{tabular}[t]{@{}r@{\hspace{0.6em}}p{\dimexpr 0.29\textwidth-7.650pt\relax-2.6em}@{}}
24 & \raggedright \rule{0pt}{2.6ex}$\langle\beta\rangle \le\mathrm{max}$ \tabularnewline
84 & \raggedright $\langle\beta\rangle \ge\mathrm{min}$\rule[-1.2ex]{0pt}{0pt} \tabularnewline
\end{tabular} \tabularnewline
\bottomrule
\end{tabular}
\end{table*}

\begin{table*}[htbp]
\centering\scriptsize
\caption{\texttt{large\_tokamak\_nof}}  %
\label{tab:problem-large-tokamak-nof}
\setlength{\aboverulesep}{0pt}\setlength{\belowrulesep}{0pt}
\begin{tabular}{@{}p{\dimexpr 0.37\textwidth-7.650pt\relax}@{\hspace{5pt}\vrule width 0.2pt\relax\hspace{5pt}}p{\dimexpr 0.15\textwidth-7.650pt\relax}@{\hspace{5pt}\vrule width 0.2pt\relax\hspace{5pt}}p{\dimexpr 0.19\textwidth-7.650pt\relax}@{\hspace{5pt}\vrule width 0.2pt\relax\hspace{5pt}}p{\dimexpr 0.29\textwidth-7.650pt\relax}@{}}
\toprule
\rule{0pt}{2.6ex}\rule[-1.2ex]{0pt}{0pt}\raggedright design variables (\texttt{ixc}) & \raggedright objective & \raggedright equality constraints (\texttt{icc}) & \raggedright inequality constraints (\texttt{icc}) \tabularnewline
\midrule
\begin{tabular}[t]{@{}r@{\hspace{0.6em}}p{\dimexpr 0.37\textwidth-7.650pt\relax-2.6em}@{}}
2 & \raggedright \rule{0pt}{2.6ex}$B_\mathrm{t}(R_0)$ $\in$ [\num{0.01}, \num{30}] T \tabularnewline
3 & \raggedright $R_0$ $\in$ [\num{8}, \num{9}] m \tabularnewline
4 & \raggedright $\langle T_\mathrm{e}\rangle$ $\in$ [\num{5.5055}, \num{100}] keV \tabularnewline
5 & \raggedright $\langle\beta\rangle$ $\in$ [\num{0.001}, \num{1}] \tabularnewline
6 & \raggedright $\langle n_\mathrm{e}\rangle$ $\in$ [\num{2e+19}, \num{1e+21}] m$^{-3}$ \tabularnewline
10 & \raggedright $H$ $\in$ [\num{0.1}, \num{1.2}] \tabularnewline
16 & \raggedright CS radial thickness $\in$ [\num{0.3}, \num{10}] m \tabularnewline
18 & \raggedright $q_{95}$ $\in$ [\num{3}, \num{50}] \tabularnewline
29 & \raggedright Bore radius $\in$ [\num{0.1}, \num{10}] m \tabularnewline
37 & \raggedright $j_\mathrm{CS,EOF}$ $\in$ [\num{100000}, \num{1e+08}] A\,m$^{-2}$ \tabularnewline
44 & \raggedright $f_\mathrm{NI}$ $\in$ [\num{0.001}, \num{1}] \tabularnewline
56 & \raggedright $\tau_\mathrm{Q}$ $\in$ [\num{0.1}, \num{100}] s \tabularnewline
57 & \raggedright TF nose case thickness $\in$ [\num{0.05}, \num{1}] m \tabularnewline
58 & \raggedright TF turn conduit thickness $\in$ [\num{0.008}, \num{0.1}] m \tabularnewline
59 & \raggedright $f_\mathrm{Cu}$ $\in$ [\num{0.5}, \num{0.94}] \tabularnewline
60 & \raggedright $I_\mathrm{turn}$ $\in$ [\num{65000}, \num{90000}] A \tabularnewline
109 & \raggedright $\langle n_{\alpha,\mathrm{th}}\rangle/\langle n_\mathrm{e}\rangle$ $\in$ [\num{0.05}, \num{0.1}] \tabularnewline
122 & \raggedright CS turn steel fraction $\in$ [\num{0.001}, \num{0.95}] \tabularnewline
135 & \raggedright $n_\mathrm{Xe}/n_\mathrm{e}$ $\in$ [\num{1e-08}, \num{0.01}] \tabularnewline
140 & \raggedright TF winding pack radial thickness $\in$ [\num{0.4}, \num{2}] m\rule[-1.2ex]{0pt}{0pt} \tabularnewline
\end{tabular} &
\begin{tabular}[t]{@{}r@{\hspace{0.6em}}p{\dimexpr 0.15\textwidth-7.650pt\relax-2.6em}@{}}
1 & \raggedright \rule{0pt}{2.6ex}min $R_0$\rule[-1.2ex]{0pt}{0pt} \tabularnewline
\end{tabular} &
\begin{tabular}[t]{@{}r@{\hspace{0.6em}}p{\dimexpr 0.19\textwidth-7.650pt\relax-2.6em}@{}}
1 & \raggedright \rule{0pt}{2.6ex}$\langle\beta\rangle$ consistency \tabularnewline
2 & \raggedright Power balance \tabularnewline
11 & \raggedright Radial build\rule[-1.2ex]{0pt}{0pt} \tabularnewline
\end{tabular} &
\begin{tabular}[t]{@{}r@{\hspace{0.6em}}p{\dimexpr 0.29\textwidth-7.650pt\relax-2.6em}@{}}
5 & \raggedright \rule{0pt}{2.6ex}$\langle n_\mathrm{e}\rangle \le n_\mathrm{lim}$ \tabularnewline
8 & \raggedright $q_\mathrm{n} \le\mathrm{max}$ \tabularnewline
9 & \raggedright $P_\mathrm{fus} \le\mathrm{max}$ \tabularnewline
13 & \raggedright $t_\mathrm{burn} \ge\mathrm{min}$ \tabularnewline
15 & \raggedright $P_\mathrm{sep} \ge P_\mathrm{LH}$ \tabularnewline
16 & \raggedright $P_\mathrm{net} \ge\mathrm{min}$ \tabularnewline
24 & \raggedright $\langle\beta\rangle \le\mathrm{max}$ \tabularnewline
25 & \raggedright $B_\mathrm{max} \le\mathrm{max}$ \tabularnewline
26 & \raggedright $j_\mathrm{CS} \le j_\mathrm{crit}$ (EOF) \tabularnewline
27 & \raggedright $j_\mathrm{CS} \le j_\mathrm{crit}$ (BOP) \tabularnewline
30 & \raggedright $P_\mathrm{aux} \le\mathrm{max}$ \tabularnewline
31 & \raggedright $\sigma_\mathrm{TF,case} \le\mathrm{max}$ \tabularnewline
32 & \raggedright $\sigma_\mathrm{TF,conduit} \le\mathrm{max}$ \tabularnewline
33 & \raggedright $j_\mathrm{TF,WP} \le j_\mathrm{crit}$ \tabularnewline
34 & \raggedright $V_\mathrm{TF} \le\mathrm{max}$ \tabularnewline
35 & \raggedright $j_\mathrm{TF,WP} \le j_\mathrm{quench}$ \tabularnewline
36 & \raggedright $\Delta T_\mathrm{TF} \ge\mathrm{min}$ \tabularnewline
60 & \raggedright $\Delta T_\mathrm{CS} \ge\mathrm{min}$ \tabularnewline
62 & \raggedright $\tau_\alpha/\tau_\mathrm{E} \ge\mathrm{min}$ \tabularnewline
65 & \raggedright $\sigma_\mathrm{VV,quench} \le\mathrm{max}$ \tabularnewline
68 & \raggedright $P_\mathrm{sep} B_\mathrm{T}/(q_{95} A R_0) \le\mathrm{max}$ \tabularnewline
72 & \raggedright $\sigma_\mathrm{CS,Tresca} \le\mathrm{max}$ \tabularnewline
81 & \raggedright $n_\mathrm{e0} \ge n_\mathrm{e,ped}$\rule[-1.2ex]{0pt}{0pt} \tabularnewline
\end{tabular} \tabularnewline
\bottomrule
\end{tabular}
\end{table*}

\begin{table*}[htbp]
\centering\scriptsize
\caption{\texttt{large\_tokamak\_eval}}  %
\label{tab:problem-large-tokamak-eval}
\setlength{\aboverulesep}{0pt}\setlength{\belowrulesep}{0pt}
\begin{tabular}{@{}p{\dimexpr 0.37\textwidth-7.650pt\relax}@{\hspace{5pt}\vrule width 0.2pt\relax\hspace{5pt}}p{\dimexpr 0.15\textwidth-7.650pt\relax}@{\hspace{5pt}\vrule width 0.2pt\relax\hspace{5pt}}p{\dimexpr 0.19\textwidth-7.650pt\relax}@{\hspace{5pt}\vrule width 0.2pt\relax\hspace{5pt}}p{\dimexpr 0.29\textwidth-7.650pt\relax}@{}}
\toprule
\rule{0pt}{2.6ex}\rule[-1.2ex]{0pt}{0pt}\raggedright design variables (\texttt{ixc}) & \raggedright objective & \raggedright equality constraints (\texttt{icc}) & \raggedright inequalities (\texttt{icc}; evaluated at the solution, not imposed) \tabularnewline
\midrule
\begin{tabular}[t]{@{}r@{\hspace{0.6em}}p{\dimexpr 0.37\textwidth-7.650pt\relax-2.6em}@{}}
4 & \raggedright \rule{0pt}{2.6ex}$\langle T_\mathrm{e}\rangle$ $\in$ [\num{5.5055}, \num{100}] keV \tabularnewline
6 & \raggedright $\langle n_\mathrm{e}\rangle$ $\in$ [\num{2e+19}, \num{1e+21}] m$^{-3}$\rule[-1.2ex]{0pt}{0pt} \tabularnewline
\end{tabular} &
\begin{tabular}[t]{@{}r@{\hspace{0.6em}}p{\dimexpr 0.15\textwidth-7.650pt\relax-2.6em}@{}}
 & \raggedright \rule{0pt}{2.6ex}none (root find)\rule[-1.2ex]{0pt}{0pt} \tabularnewline
\end{tabular} &
\begin{tabular}[t]{@{}r@{\hspace{0.6em}}p{\dimexpr 0.19\textwidth-7.650pt\relax-2.6em}@{}}
1 & \raggedright \rule{0pt}{2.6ex}$\langle\beta\rangle$ consistency \tabularnewline
2 & \raggedright Power balance\rule[-1.2ex]{0pt}{0pt} \tabularnewline
\end{tabular} &
\begin{tabular}[t]{@{}r@{\hspace{0.6em}}p{\dimexpr 0.29\textwidth-7.650pt\relax-2.6em}@{}}
5 & \raggedright \rule{0pt}{2.6ex}$\langle n_\mathrm{e}\rangle \le n_\mathrm{lim}$ \tabularnewline
8 & \raggedright $q_\mathrm{n} \le\mathrm{max}$ \tabularnewline
9 & \raggedright $P_\mathrm{fus} \le\mathrm{max}$ \tabularnewline
13 & \raggedright $t_\mathrm{burn} \ge\mathrm{min}$ \tabularnewline
15 & \raggedright $P_\mathrm{sep} \ge P_\mathrm{LH}$ \tabularnewline
16 & \raggedright $P_\mathrm{net} \ge\mathrm{min}$ \tabularnewline
24 & \raggedright $\langle\beta\rangle \le\mathrm{max}$ \tabularnewline
25 & \raggedright $B_\mathrm{max} \le\mathrm{max}$ \tabularnewline
26 & \raggedright $j_\mathrm{CS} \le j_\mathrm{crit}$ (EOF) \tabularnewline
27 & \raggedright $j_\mathrm{CS} \le j_\mathrm{crit}$ (BOP) \tabularnewline
30 & \raggedright $P_\mathrm{aux} \le\mathrm{max}$ \tabularnewline
31 & \raggedright $\sigma_\mathrm{TF,case} \le\mathrm{max}$ \tabularnewline
32 & \raggedright $\sigma_\mathrm{TF,conduit} \le\mathrm{max}$ \tabularnewline
33 & \raggedright $j_\mathrm{TF,WP} \le j_\mathrm{crit}$ \tabularnewline
34 & \raggedright $V_\mathrm{TF} \le\mathrm{max}$ \tabularnewline
35 & \raggedright $j_\mathrm{TF,WP} \le j_\mathrm{quench}$ \tabularnewline
36 & \raggedright $\Delta T_\mathrm{TF} \ge\mathrm{min}$ \tabularnewline
60 & \raggedright $\Delta T_\mathrm{CS} \ge\mathrm{min}$ \tabularnewline
62 & \raggedright $\tau_\alpha/\tau_\mathrm{E} \ge\mathrm{min}$ \tabularnewline
65 & \raggedright $\sigma_\mathrm{VV,quench} \le\mathrm{max}$ \tabularnewline
68 & \raggedright $P_\mathrm{sep} B_\mathrm{T}/(q_{95} A R_0) \le\mathrm{max}$ \tabularnewline
72 & \raggedright $\sigma_\mathrm{CS,Tresca} \le\mathrm{max}$ \tabularnewline
81 & \raggedright $n_\mathrm{e0} \ge n_\mathrm{e,ped}$\rule[-1.2ex]{0pt}{0pt} \tabularnewline
\end{tabular} \tabularnewline
\bottomrule
\end{tabular}
\end{table*}

\begin{table*}[htbp]
\centering\scriptsize
\caption{\texttt{low\_aspect\_ratio\_DEMO}}  %
\label{tab:problem-low-aspect-ratio-DEMO}
\setlength{\aboverulesep}{0pt}\setlength{\belowrulesep}{0pt}
\begin{tabular}{@{}p{\dimexpr 0.37\textwidth-7.650pt\relax}@{\hspace{5pt}\vrule width 0.2pt\relax\hspace{5pt}}p{\dimexpr 0.15\textwidth-7.650pt\relax}@{\hspace{5pt}\vrule width 0.2pt\relax\hspace{5pt}}p{\dimexpr 0.19\textwidth-7.650pt\relax}@{\hspace{5pt}\vrule width 0.2pt\relax\hspace{5pt}}p{\dimexpr 0.29\textwidth-7.650pt\relax}@{}}
\toprule
\rule{0pt}{2.6ex}\rule[-1.2ex]{0pt}{0pt}\raggedright design variables (\texttt{ixc}) & \raggedright objective & \raggedright equality constraints (\texttt{icc}) & \raggedright inequality constraints (\texttt{icc}) \tabularnewline
\midrule
\begin{tabular}[t]{@{}r@{\hspace{0.6em}}p{\dimexpr 0.37\textwidth-7.650pt\relax-2.6em}@{}}
2 & \raggedright \rule{0pt}{2.6ex}$B_\mathrm{t}(R_0)$ $\in$ [\num{0.01}, \num{20}] T \tabularnewline
3 & \raggedright $R_0$ $\in$ [\num{8}, \num{8.6}] m \tabularnewline
4 & \raggedright $\langle T_\mathrm{e}\rangle$ $\in$ [\num{5}, \num{150}] keV \tabularnewline
5 & \raggedright $\langle\beta\rangle$ $\in$ [\num{0.001}, \num{1}] \tabularnewline
6 & \raggedright $\langle n_\mathrm{e}\rangle$ $\in$ [\num{2e+19}, \num{1e+21}] m$^{-3}$ \tabularnewline
13 & \raggedright TF inboard leg thickness $\in$ [\num{0.3}, \num{5}] m \tabularnewline
16 & \raggedright CS radial thickness $\in$ [\num{0.1}, \num{10}] m \tabularnewline
18 & \raggedright $q_{95}$ $\in$ [\num{3.6}, \num{50}] \tabularnewline
29 & \raggedright Bore radius $\in$ [\num{0.1}, \num{10}] m \tabularnewline
37 & \raggedright $j_\mathrm{CS,EOF}$ $\in$ [\num{100000}, \num{1e+08}] A\,m$^{-2}$ \tabularnewline
41 & \raggedright $j_\mathrm{CS,BOP}/j_\mathrm{CS,EOF}$ $\in$ [\num{0.001}, \num{1}] \tabularnewline
44 & \raggedright $f_\mathrm{NI}$ $\in$ [\num{0.001}, \num{1}] \tabularnewline
56 & \raggedright $\tau_\mathrm{Q}$ $\in$ [\num{0.1}, \num{100}] s \tabularnewline
57 & \raggedright TF nose case thickness $\in$ [\num{0.05}, \num{1}] m \tabularnewline
58 & \raggedright TF turn conduit thickness $\in$ [\num{0.008}, \num{0.1}] m \tabularnewline
59 & \raggedright $f_\mathrm{Cu}$ $\in$ [\num{0.5}, \num{0.94}] \tabularnewline
109 & \raggedright $\langle n_{\alpha,\mathrm{th}}\rangle/\langle n_\mathrm{e}\rangle$ $\in$ [\num{0.05}, \num{0.15}] \tabularnewline
122 & \raggedright CS turn steel fraction $\in$ [\num{0.001}, \num{0.95}] \tabularnewline
135 & \raggedright $n_\mathrm{Xe}/n_\mathrm{e}$ $\in$ [\num{1e-08}, \num{0.01}]\rule[-1.2ex]{0pt}{0pt} \tabularnewline
\end{tabular} &
\begin{tabular}[t]{@{}r@{\hspace{0.6em}}p{\dimexpr 0.15\textwidth-7.650pt\relax-2.6em}@{}}
-14 & \raggedright \rule{0pt}{2.6ex}max $t_\mathrm{burn}$\rule[-1.2ex]{0pt}{0pt} \tabularnewline
\end{tabular} &
\begin{tabular}[t]{@{}r@{\hspace{0.6em}}p{\dimexpr 0.19\textwidth-7.650pt\relax-2.6em}@{}}
1 & \raggedright \rule{0pt}{2.6ex}$\langle\beta\rangle$ consistency \tabularnewline
2 & \raggedright Power balance \tabularnewline
11 & \raggedright Radial build \tabularnewline
30 & \raggedright $P_\mathrm{aux} \le\mathrm{max}$\rule[-1.2ex]{0pt}{0pt} \tabularnewline
\end{tabular} &
\begin{tabular}[t]{@{}r@{\hspace{0.6em}}p{\dimexpr 0.29\textwidth-7.650pt\relax-2.6em}@{}}
5 & \raggedright \rule{0pt}{2.6ex}$\langle n_\mathrm{e}\rangle \le n_\mathrm{lim}$ \tabularnewline
8 & \raggedright $q_\mathrm{n} \le\mathrm{max}$ \tabularnewline
13 & \raggedright $t_\mathrm{burn} \ge\mathrm{min}$ \tabularnewline
15 & \raggedright $P_\mathrm{sep} \ge P_\mathrm{LH}$ \tabularnewline
16 & \raggedright $P_\mathrm{net} \ge\mathrm{min}$ \tabularnewline
24 & \raggedright $\langle\beta\rangle \le\mathrm{max}$ \tabularnewline
26 & \raggedright $j_\mathrm{CS} \le j_\mathrm{crit}$ (EOF) \tabularnewline
27 & \raggedright $j_\mathrm{CS} \le j_\mathrm{crit}$ (BOP) \tabularnewline
31 & \raggedright $\sigma_\mathrm{TF,case} \le\mathrm{max}$ \tabularnewline
32 & \raggedright $\sigma_\mathrm{TF,conduit} \le\mathrm{max}$ \tabularnewline
33 & \raggedright $j_\mathrm{TF,WP} \le j_\mathrm{crit}$ \tabularnewline
34 & \raggedright $V_\mathrm{TF} \le\mathrm{max}$ \tabularnewline
35 & \raggedright $j_\mathrm{TF,WP} \le j_\mathrm{quench}$ \tabularnewline
36 & \raggedright $\Delta T_\mathrm{TF} \ge\mathrm{min}$ \tabularnewline
60 & \raggedright $\Delta T_\mathrm{CS} \ge\mathrm{min}$ \tabularnewline
62 & \raggedright $\tau_\alpha/\tau_\mathrm{E} \ge\mathrm{min}$ \tabularnewline
65 & \raggedright $\sigma_\mathrm{VV,quench} \le\mathrm{max}$ \tabularnewline
68 & \raggedright $P_\mathrm{sep} B_\mathrm{T}/(q_{95} A R_0) \le\mathrm{max}$ \tabularnewline
72 & \raggedright $\sigma_\mathrm{CS,Tresca} \le\mathrm{max}$ \tabularnewline
81 & \raggedright $n_\mathrm{e0} \ge n_\mathrm{e,ped}$ \tabularnewline
90 & \raggedright $N_\mathrm{cycles,CS} \ge\mathrm{min}$\rule[-1.2ex]{0pt}{0pt} \tabularnewline
\end{tabular} \tabularnewline
\bottomrule
\end{tabular}
\end{table*}

\begin{table*}[htbp]
\centering\scriptsize
\caption{\texttt{spherical\_tokamak\_eval}}  %
\label{tab:problem-spherical-tokamak-eval}
\setlength{\aboverulesep}{0pt}\setlength{\belowrulesep}{0pt}
\begin{tabular}{@{}p{\dimexpr 0.37\textwidth-7.650pt\relax}@{\hspace{5pt}\vrule width 0.2pt\relax\hspace{5pt}}p{\dimexpr 0.15\textwidth-7.650pt\relax}@{\hspace{5pt}\vrule width 0.2pt\relax\hspace{5pt}}p{\dimexpr 0.19\textwidth-7.650pt\relax}@{\hspace{5pt}\vrule width 0.2pt\relax\hspace{5pt}}p{\dimexpr 0.29\textwidth-7.650pt\relax}@{}}
\toprule
\rule{0pt}{2.6ex}\rule[-1.2ex]{0pt}{0pt}\raggedright design variables (\texttt{ixc}) & \raggedright objective & \raggedright equality constraints (\texttt{icc}) & \raggedright inequalities (\texttt{icc}; evaluated at the solution, not imposed) \tabularnewline
\midrule
\begin{tabular}[t]{@{}r@{\hspace{0.6em}}p{\dimexpr 0.37\textwidth-7.650pt\relax-2.6em}@{}}
4 & \raggedright \rule{0pt}{2.6ex}$\langle T_\mathrm{e}\rangle$ $\in$ [\num{5}, \num{25}] keV \tabularnewline
6 & \raggedright $\langle n_\mathrm{e}\rangle$ $\in$ [\num{5e+19}, \num{5e+20}] m$^{-3}$ \tabularnewline
29 & \raggedright Bore radius $\in$ [\num{0.1}, \num{0.8}] m\rule[-1.2ex]{0pt}{0pt} \tabularnewline
\end{tabular} &
\begin{tabular}[t]{@{}r@{\hspace{0.6em}}p{\dimexpr 0.15\textwidth-7.650pt\relax-2.6em}@{}}
 & \raggedright \rule{0pt}{2.6ex}none (root find)\rule[-1.2ex]{0pt}{0pt} \tabularnewline
\end{tabular} &
\begin{tabular}[t]{@{}r@{\hspace{0.6em}}p{\dimexpr 0.19\textwidth-7.650pt\relax-2.6em}@{}}
1 & \raggedright \rule{0pt}{2.6ex}$\langle\beta\rangle$ consistency \tabularnewline
2 & \raggedright Power balance \tabularnewline
11 & \raggedright Radial build\rule[-1.2ex]{0pt}{0pt} \tabularnewline
\end{tabular} &
\begin{tabular}[t]{@{}r@{\hspace{0.6em}}p{\dimexpr 0.29\textwidth-7.650pt\relax-2.6em}@{}}
5 & \raggedright \rule{0pt}{2.6ex}$\langle n_\mathrm{e}\rangle \le n_\mathrm{lim}$ \tabularnewline
9 & \raggedright $P_\mathrm{fus} \le\mathrm{max}$ \tabularnewline
15 & \raggedright $P_\mathrm{sep} \ge P_\mathrm{LH}$ \tabularnewline
16 & \raggedright $P_\mathrm{net} \ge\mathrm{min}$ \tabularnewline
17 & \raggedright $f_\mathrm{rad} \le\mathrm{max}$ \tabularnewline
24 & \raggedright $\langle\beta\rangle \le\mathrm{max}$ \tabularnewline
30 & \raggedright $P_\mathrm{aux} \le\mathrm{max}$ \tabularnewline
31 & \raggedright $\sigma_\mathrm{TF,case} \le\mathrm{max}$ \tabularnewline
32 & \raggedright $\sigma_\mathrm{TF,conduit} \le\mathrm{max}$ \tabularnewline
33 & \raggedright $j_\mathrm{TF,WP} \le j_\mathrm{crit}$ \tabularnewline
46 & \raggedright $I_\mathrm{p}/I_\mathrm{rod} \le\mathrm{max}$ \tabularnewline
56 & \raggedright $P_\mathrm{sep}/R_0 \le\mathrm{max}$ \tabularnewline
62 & \raggedright $\tau_\alpha/\tau_\mathrm{E} \ge\mathrm{min}$ \tabularnewline
67 & \raggedright $q_\mathrm{FW,rad} \le\mathrm{max}$ \tabularnewline
81 & \raggedright $n_\mathrm{e0} \ge n_\mathrm{e,ped}$\rule[-1.2ex]{0pt}{0pt} \tabularnewline
\end{tabular} \tabularnewline
\bottomrule
\end{tabular}
\end{table*}

\begin{table*}[htbp]
\centering\scriptsize
\caption{\texttt{st\_regression}}  %
\label{tab:problem-st-regression}
\setlength{\aboverulesep}{0pt}\setlength{\belowrulesep}{0pt}
\begin{tabular}{@{}p{\dimexpr 0.37\textwidth-7.650pt\relax}@{\hspace{5pt}\vrule width 0.2pt\relax\hspace{5pt}}p{\dimexpr 0.15\textwidth-7.650pt\relax}@{\hspace{5pt}\vrule width 0.2pt\relax\hspace{5pt}}p{\dimexpr 0.19\textwidth-7.650pt\relax}@{\hspace{5pt}\vrule width 0.2pt\relax\hspace{5pt}}p{\dimexpr 0.29\textwidth-7.650pt\relax}@{}}
\toprule
\rule{0pt}{2.6ex}\rule[-1.2ex]{0pt}{0pt}\raggedright design variables (\texttt{ixc}) & \raggedright objective & \raggedright equality constraints (\texttt{icc}) & \raggedright inequality constraints (\texttt{icc}) \tabularnewline
\midrule
\begin{tabular}[t]{@{}r@{\hspace{0.6em}}p{\dimexpr 0.37\textwidth-7.650pt\relax-2.6em}@{}}
4 & \raggedright \rule{0pt}{2.6ex}$\langle T_\mathrm{e}\rangle$ $\in$ [\num{5}, \num{25}] keV \tabularnewline
5 & \raggedright $\langle\beta\rangle$ $\in$ [\num{0.01}, \num{1}] \tabularnewline
6 & \raggedright $\langle n_\mathrm{e}\rangle$ $\in$ [\num{5e+19}, \num{5e+20}] m$^{-3}$ \tabularnewline
10 & \raggedright $H$ $\in$ [\num{0.8}, \num{1.2}] \tabularnewline
16 & \raggedright CS radial thickness $\in$ [\num{0.03}, \num{0.5}] m \tabularnewline
18 & \raggedright $q_{95}$ $\in$ [\num{3}, \num{20}] \tabularnewline
29 & \raggedright Bore radius $\in$ [\num{0.1}, \num{0.8}] m \tabularnewline
57 & \raggedright TF nose case thickness $\in$ [\num{0.01}, \num{0.5}] m \tabularnewline
93 & \raggedright Inboard shield thickness $\in$ [\num{0.01}, \num{1}] m \tabularnewline
109 & \raggedright $\langle n_{\alpha,\mathrm{th}}\rangle/\langle n_\mathrm{e}\rangle$ $\in$ [\num{0.001}, \num{1}] \tabularnewline
135 & \raggedright $n_\mathrm{Xe}/n_\mathrm{e}$ $\in$ [\num{1e-08}, \num{0.01}] \tabularnewline
140 & \raggedright TF winding pack radial thickness $\in$ [\num{0.4}, \num{0.8}] m \tabularnewline
145 & \raggedright $f_\mathrm{GW,ped}$ $\in$ [\num{0.1}, \num{1}] \tabularnewline
152 & \raggedright $f_\mathrm{GW,sep}$ $\in$ [\num{0.001}, \num{0.1}]\rule[-1.2ex]{0pt}{0pt} \tabularnewline
\end{tabular} &
\begin{tabular}[t]{@{}r@{\hspace{0.6em}}p{\dimexpr 0.15\textwidth-7.650pt\relax-2.6em}@{}}
-5 & \raggedright \rule{0pt}{2.6ex}max $Q$\rule[-1.2ex]{0pt}{0pt} \tabularnewline
\end{tabular} &
\begin{tabular}[t]{@{}r@{\hspace{0.6em}}p{\dimexpr 0.19\textwidth-7.650pt\relax-2.6em}@{}}
1 & \raggedright \rule{0pt}{2.6ex}$\langle\beta\rangle$ consistency \tabularnewline
2 & \raggedright Power balance \tabularnewline
11 & \raggedright Radial build\rule[-1.2ex]{0pt}{0pt} \tabularnewline
\end{tabular} &
\begin{tabular}[t]{@{}r@{\hspace{0.6em}}p{\dimexpr 0.29\textwidth-7.650pt\relax-2.6em}@{}}
5 & \raggedright \rule{0pt}{2.6ex}$\langle n_\mathrm{e}\rangle \le n_\mathrm{lim}$ \tabularnewline
9 & \raggedright $P_\mathrm{fus} \le\mathrm{max}$ \tabularnewline
15 & \raggedright $P_\mathrm{sep} \ge P_\mathrm{LH}$ \tabularnewline
16 & \raggedright $P_\mathrm{net} \ge\mathrm{min}$ \tabularnewline
17 & \raggedright $f_\mathrm{rad} \le\mathrm{max}$ \tabularnewline
24 & \raggedright $\langle\beta\rangle \le\mathrm{max}$ \tabularnewline
30 & \raggedright $P_\mathrm{aux} \le\mathrm{max}$ \tabularnewline
31 & \raggedright $\sigma_\mathrm{TF,case} \le\mathrm{max}$ \tabularnewline
32 & \raggedright $\sigma_\mathrm{TF,conduit} \le\mathrm{max}$ \tabularnewline
33 & \raggedright $j_\mathrm{TF,WP} \le j_\mathrm{crit}$ \tabularnewline
46 & \raggedright $I_\mathrm{p}/I_\mathrm{rod} \le\mathrm{max}$ \tabularnewline
56 & \raggedright $P_\mathrm{sep}/R_0 \le\mathrm{max}$ \tabularnewline
62 & \raggedright $\tau_\alpha/\tau_\mathrm{E} \ge\mathrm{min}$ \tabularnewline
67 & \raggedright $q_\mathrm{FW,rad} \le\mathrm{max}$ \tabularnewline
81 & \raggedright $n_\mathrm{e0} \ge n_\mathrm{e,ped}$\rule[-1.2ex]{0pt}{0pt} \tabularnewline
\end{tabular} \tabularnewline
\bottomrule
\end{tabular}
\end{table*}

\begin{table*}[htbp]
\centering\scriptsize\setlength{\tabcolsep}{2.5pt}
\caption{Integer switches used as in PROCESS for the seven configurations}  %
\label{tab:problem-switches}
\begin{tabular}{@{}lccccccc@{}}
\toprule
switch & \rotatebox{70}{\texttt{stellarator\_helias}} & \rotatebox{70}{\texttt{helias\_5b}} & \rotatebox{70}{\texttt{large\_tokamak\_nof}} & \rotatebox{70}{\texttt{large\_tokamak\_eval}} & \rotatebox{70}{\texttt{low\_aspect\_ratio\_DEMO}} & \rotatebox{70}{\texttt{spherical\_tokamak\_eval}} & \rotatebox{70}{\texttt{st\_regression}} \\
\midrule
\texttt{i\_\allowbreak{}blkt\_\allowbreak{}inboard} & 1 & 1 & 1 & 1 & 1 & 0 & 0 \\
\texttt{i\_\allowbreak{}cs\_\allowbreak{}precomp} & 1 & 1 & 1 & 1 & 1 & 0 & 0 \\
\texttt{i\_\allowbreak{}r\_\allowbreak{}cp\_\allowbreak{}top} & 0 & 0 & 0 & 0 & 0 & 2 & 2 \\
\texttt{iohcl} & 1 & 1 & 1 & 1 & 1 & 0 & 0 \\
\texttt{i\_\allowbreak{}q95\_\allowbreak{}fixed} & 0 & 0 & 0 & 0 & 1 & 0 & 0 \\
\texttt{ibkt\_\allowbreak{}life} & 0 & 0 & 0 & 0 & 1 & 0 & 0 \\
\texttt{ifueltyp} & 0 & 0 & 0 & 0 & 1 & 0 & 0 \\
\texttt{lsa} & 2 & 2 & 4 & 4 & 2 & 2 & 2 \\
\texttt{output\_\allowbreak{}costs} & 1 & 1 & 1 & 1 & 0 & 1 & 1 \\
\texttt{i\_\allowbreak{}hcd\_\allowbreak{}primary} & 5 & 5 & 10 & 10 & 10 & 13 & 13 \\
\texttt{i\_\allowbreak{}fw\_\allowbreak{}blkt\_\allowbreak{}vv\_\allowbreak{}shape} & 2 & 2 & 2 & 2 & 2 & 1 & 1 \\
\texttt{i\_\allowbreak{}p\_\allowbreak{}coolant\_\allowbreak{}pumping} & 1 & 0 & 3 & 3 & 3 & 3 & 3 \\
\texttt{inuclear} & 0 & 0 & 1 & 1 & 1 & 0 & 0 \\
\texttt{ipowerflow} & 1 & 1 & 0 & 0 & 0 & 0 & 0 \\
\texttt{i\_\allowbreak{}cs\_\allowbreak{}superconductor} & 1 & 1 & 1 & 1 & 5 & 1 & 1 \\
\texttt{i\_\allowbreak{}pf\_\allowbreak{}superconductor} & 1 & 1 & 3 & 3 & 3 & 9 & 9 \\
\texttt{i\_\allowbreak{}r\_\allowbreak{}pf\_\allowbreak{}outside\_\allowbreak{}tf\_\allowbreak{}placement} & 0 & 0 & 0 & 0 & 0 & 1 & 1 \\
\texttt{i\_\allowbreak{}alphaj} & 0 & 0 & 0 & 1 & 0 & 0 & 0 \\
\texttt{i\_\allowbreak{}beta\_\allowbreak{}component} & 0 & 0 & 1 & 1 & 1 & 3 & 3 \\
\texttt{i\_\allowbreak{}beta\_\allowbreak{}norm\_\allowbreak{}max} & 1 & 1 & 1 & 1 & 1 & 0 & 0 \\
\texttt{i\_\allowbreak{}bootstrap\_\allowbreak{}current} & 3 & 3 & 4 & 4 & 4 & 4 & 4 \\
\texttt{i\_\allowbreak{}confinement\_\allowbreak{}time} & 38 & 38 & 34 & 34 & 34 & 34 & 34 \\
\texttt{i\_\allowbreak{}density\_\allowbreak{}limit} & 8 & 8 & 7 & 7 & 7 & 7 & 7 \\
\texttt{i\_\allowbreak{}diamagnetic\_\allowbreak{}current} & 0 & 0 & 0 & 0 & 0 & 2 & 2 \\
\texttt{i\_\allowbreak{}ind\_\allowbreak{}plasma\_\allowbreak{}internal\_\allowbreak{}norm} & 0 & 0 & 0 & 1 & 0 & 0 & 0 \\
\texttt{i\_\allowbreak{}pfirsch\_\allowbreak{}schluter\_\allowbreak{}current} & 0 & 0 & 0 & 0 & 0 & 1 & 1 \\
\texttt{i\_\allowbreak{}plasma\_\allowbreak{}current} & 4 & 4 & 4 & 4 & 4 & 9 & 9 \\
\texttt{i\_\allowbreak{}plasma\_\allowbreak{}geometry} & 0 & 0 & 0 & 0 & 10 & 0 & 0 \\
\texttt{i\_\allowbreak{}plasma\_\allowbreak{}ignited} & 1 & 1 & 0 & 0 & 0 & 0 & 0 \\
\texttt{i\_\allowbreak{}plasma\_\allowbreak{}pedestal} & 0 & 0 & 1 & 1 & 1 & 1 & 1 \\
\texttt{i\_\allowbreak{}single\_\allowbreak{}null} & 1 & 1 & 1 & 1 & 1 & 0 & 0 \\
\texttt{itart} & 0 & 0 & 0 & 0 & 0 & 1 & 1 \\
\texttt{itartpf} & 0 & 0 & 0 & 0 & 0 & 1 & 1 \\
\texttt{i\_\allowbreak{}pulsed\_\allowbreak{}plant} & 0 & 0 & 1 & 1 & 1 & 0 & 0 \\
\texttt{istell} & 6 & 1 & 0 & 0 & 0 & 0 & 0 \\
\texttt{i\_\allowbreak{}tf\_\allowbreak{}turn\_\allowbreak{}type} & 1 & 1 & 1 & 1 & 1 & 2 & 2 \\
\texttt{i\_\allowbreak{}cp\_\allowbreak{}joints} & -1 & -1 & -1 & -1 & -1 & 0 & 0 \\
\texttt{i\_\allowbreak{}tf\_\allowbreak{}bucking} & -1 & -1 & -1 & -1 & -1 & 1 & 1 \\
\texttt{i\_\allowbreak{}tf\_\allowbreak{}sc\_\allowbreak{}mat} & 1 & 1 & 1 & 1 & 5 & 9 & 9 \\
\texttt{i\_\allowbreak{}tf\_\allowbreak{}shape} & 0 & 0 & 0 & 0 & 0 & 2 & 2 \\
\texttt{i\_\allowbreak{}tf\_\allowbreak{}stress\_\allowbreak{}model} & 1 & 1 & 1 & 1 & 1 & 0 & 0 \\
\texttt{i\_\allowbreak{}tf\_\allowbreak{}turns\_\allowbreak{}integer} & 0 & 0 & 0 & 0 & 1 & 0 & 0 \\
\texttt{i\_\allowbreak{}tf\_\allowbreak{}wp\_\allowbreak{}geom} & -1 & -1 & -1 & -1 & -1 & 2 & 2 \\
\texttt{pulsetimings} & 1 & 1 & 0 & 0 & 0 & 1 & 1 \\
\bottomrule
\end{tabular}
\end{table*}

\clearpage %

\bibliography{references,zoteroRefs}

\end{document}